# Light-Matter Interaction: Conversion of Optical Energy and Momentum to Mechanical Vibrations and Phonons

Masud Mansuripur, College of Optical Sciences, The University of Arizona, Tucson

**Abstract**. Reflection, refraction, and absorption of light by material media are, in general, accompanied by a transfer of optical energy and momentum to the media. Consequently, the eigen-modes of mechanical vibration (phonons) created in the process must distribute the acquired energy and momentum throughout the material medium. However, unlike photons, phonons do not carry momentum. What happens to the material medium in its interactions with light, therefore, requires careful consideration if the conservation laws are to be upheld. The present paper addresses some of the mechanisms by which the electromagnetic momentum of light is carried away by mechanical vibrations.

**1. Introduction**. When a light pulse arrives at the surface of a mirror, it bounces back and transfers twice its initial momentum to the mirror.[1] The mechanical motion thus set off at the front facet of the mirror travels forward and gives rise to elastic vibrations through the entire thickness of the mirror and its substrate.[2,3] Similarly, the passage of a light pulse through a transparent dielectric slab sets in motion elastic waves of mechanical vibration, which mediate the exchange of momentum between the optical wave and the material medium of the slab.[4-11] The goal of the present paper is to analyze the aforementioned problems and to derive expressions for the elastic waves excited by the electromagnetic (EM) fields, emphasizing in particular the energy and momentum associated with these elastic vibrations.

In preparation for the analysis, we begin in Sec. 2 with an elementary treatment of the continuum mechanics of solid objects, including the properties of elastic waves excited within these bodies. Several examples illustrate the one-dimensional motion of a solid object in the presence of external excitations, when the object's boundaries are fixed, and also when one or both of its boundaries are free. Then, in Sec. 3, the results of Sec. 2 pertaining to longitudinal elastic waves are extended to the case of transverse waves, where the conservation of angular momentum introduces additional complications, which require extra care and attention. Section 4 is devoted to an analysis of systems in a steady-state of motion, after the initial vibrations have died down, and the effects of external forces are reduced to static deformations of the solid object experiencing a constant and uniform acceleration.

In Sec. 5 we derive the general equation of motion for a flexible one-dimensional rod that is free to move in three-dimensional space while undergoing arbitrary elastic deformations. The motion of the rod under these circumstances is quite complicated, and the corresponding equation is not amenable to analytic solutions for any specific examples. However, we will show the consistency of the equation of motion with the conservation laws of energy and linear as well as angular momentum. In Sec. 6 we return to the simpler problems associated with longitudinal vibrations of elastic rods in one dimension, and show the effects of a light pulse impinging on a high-reflectivity metallic mirror, as well as those of a light pulse entering a transparent rod from one end (with and without anti-reflection coating), then propagating along the length of the rod.

Elastic vibrations of one-dimensional inhomogeneous media are taken up in Sec. 7. Here we derive general expressions for acoustic and optical phonons in periodic structures, and discuss the corresponding dispersion relations as well as issues related to energy and momentum conservation. Also discussed are the extension of these methods to more general situations using the theory of eigen-functions and eigen-values. Some concluding remarks appear in Sec. 8. The three appendices describe the Fourier transform theorems used throughout the paper, the optics of plane EM waves in simple optical systems, and technical details related to the topic of Sec. 7.

**2. Continuum mechanics of acoustic vibrations and phonons.** This section describes the basic equation of motion and its solution for longitudinal vibrations of a homogeneous, one-dimensional medium modelled as a string of point-particles connected by short, identical springs. With reference to Fig.1, let $u(x,t)$ be the displacement from the equilibrium position at location $x$ and time $t$, $\mu$ the mass-density (i.e., mass per unit-length along the $x$-axis), and $\sigma$ the spring constant (or stiffness coefficient). Newton's second law of motion may be written as follows:

$$\sigma\left[\frac{u(x+\Delta x) - u(x)}{\Delta x}\right] - \sigma\left[\frac{u(x) - u(x-\Delta x)}{\Delta x}\right] = (\mu\Delta x)\partial_t^2 u(x,t)$$

$$\rightarrow \quad (\sigma/\mu)\partial_x^2 u(x,t) = \partial_t^2 u(x,t). \tag{1}$$

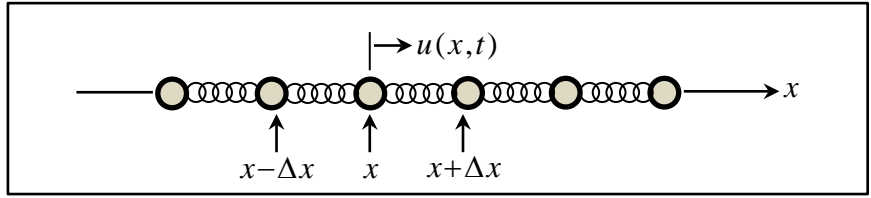

**Fig.1.** Approximating a continuous, one-dimensional slab of homogeneous material with a linear chain of point-particles connected via short springs of length $\Delta x$. The equilibrium position of the $n^{\text{th}}$ particle is $x = n\Delta x$, and the particle's displacement from equilibrium at location $x$ and time $t$ is denoted by $u(x,t)$. The mass-density of the continuum being $\mu$, each point-particle has mass $\mu\Delta x$. Denoting the spring constant by $\sigma$, the force exerted by each spring on its adjacent particles is given by $\pm\sigma\partial_x u(x,t)$.

Defining the phase-velocity of wave propagation along the $x$-axis by $v_p = \sqrt{\sigma/\mu}$, we may solve Eq.(1) by the method of separation of variables, where $u(x,t)$ is written as $f(x)g(t)$, and the separated equations become $\partial_t^2 g(t) = -\omega^2 g(t)$ and $\partial_x^2 f(x) = -(\omega/v_p)^2 f(x)$. Here we have defined the real-valued constant $\omega$ as the (arbitrary) frequency of vibrations in the time domain. Similarly, the real-valued constant $k = \omega/v_p$ is the frequency of vibrations in the space domain. The general form of the separable solution of Eq.(1) is thus written

$$u_\pm(x,t) = U_\pm(k)\exp[i(kx \pm \omega t)]. \tag{2}$$

The initial state of the medium at $t = 0$, that is, its position $u(x, t = 0)$ and its local velocity profile $v(x, t = 0) = \partial_t u(x, t = 0)$ may thus be expressed as follows:

$$u(x, t = 0) = \frac{1}{2\pi}\int_{-\infty}^{\infty}[U_+(k) + U_-(k)]\exp(ikx)\,dk, \tag{3a}$$

$$v(x, t = 0) = \frac{v_p}{2\pi}\int_{-\infty}^{\infty}ik[U_+(k) - U_-(k)]\exp(ikx)\,dk. \tag{3b}$$

The functions $U_\pm(k)$ are readily derived from the Fourier-transformed initial conditions $U_0(k)$ and $V_0(k)$. In particular, if $u(x, t = 0) = 0$, we will have $U_+(k) = -U_-(k) = V_0(k)/(2ikv_p)$. In the absence of dispersion, that is, when $v_p$ is independent of the frequency $\omega$, the general solution of Eq.(1) will be $u(x,t) = u_+(x + v_p t) + u_-(x - v_p t)$. The velocity profile will then be $v(x,t) = v_p \partial_x[u_+(x + v_p t) - u_-(x - v_p t)]$.

For a uniform, homogeneous, dispersionless medium that is also infinitely-long along the propagation direction $x$, one can prove the conservation of energy (kinetic $\mathcal{E}_K$ plus potential $\mathcal{E}_P$) and of linear momentum $p_x$, as follows:

$$\mathcal{E} = \mathcal{E}_K + \mathcal{E}_P = \tfrac{1}{2}\mu \int_{-\infty}^{\infty} v^2(x,t) dx + \tfrac{1}{2}\sigma \int_{-\infty}^{\infty} [\partial_x u(x,t)]^2 dx$$

$$= \tfrac{1}{2}\mu v_p^2 \int_{-\infty}^{\infty} [\partial_x u_+(x + v_p t) - \partial_x u_-(x - v_p t)]^2 dx$$

$$+ \tfrac{1}{2}\sigma \int_{-\infty}^{\infty} [\partial_x u_+(x + v_p t) + \partial_x u_-(x - v_p t)]^2 dx$$

$$= \sigma \int_{-\infty}^{\infty} \{[\partial_x u_+(x)]^2 + [\partial_x u_-(x)]^2\} dx. \tag{4}$$

$$p_x = \int_{-\infty}^{\infty} \mu v(x,t) dx = \mu v_p \int_{-\infty}^{\infty} \partial_x [u_+(x + v_p t) - u_-(x - v_p t)] dx$$

$$= \mu v_p [u_+(\infty) - u_-(\infty) - u_+(-\infty) + u_-(-\infty)]. \tag{5}$$

Both $\mathcal{E}$ and $p_x$ are thus seen to be time-independent, which is proof that the total energy and linear momentum of the system are conserved. Conservation of angular momentum is trivial to prove in the present case of longitudinal vibrations along $x$; just pick an arbitrary reference point $x_0$ on the $x$-axis and observe that $\boldsymbol{L} = \int_{-\infty}^{\infty} (x - x_0)\hat{\boldsymbol{x}} \times \mu v(x,t)\hat{\boldsymbol{x}} \, dx$ is zero at all times.

**Example 1**. Suppose that the initial position and velocity of our one-dimensional medium are specified as $u(x, t = 0) = 0$ and $v(x, t = 0) = v_0/[1 + (x/w_0)^2]$, respectively. As shown in Appendix A, the Fourier transform of the initial velocity will be $V_0(k) = \pi v_0 w_0 \exp(-w_0|k|)$ and, consequently, $U_\pm(k) = \pm \pi w_0 v_0 \exp(-w_0|k|)/(2ikv_p)$, which leads to $u_\pm(x, 0) = \pm\tfrac{1}{2}(w_0 v_0/v_p)\arctan(x/w_0)$, as depicted in Fig.2. The temporal evolution of our system having the postulated initial conditions may thus be written as follows:

$$u(x,t) = u_+(x,t) + u_-(x,t) = \frac{w_0 v_0}{2v_p} \int_{-\infty}^{\infty} k^{-1} \exp(-w_0|k| + ikx) \sin(\omega t) \, dk. \tag{6}$$

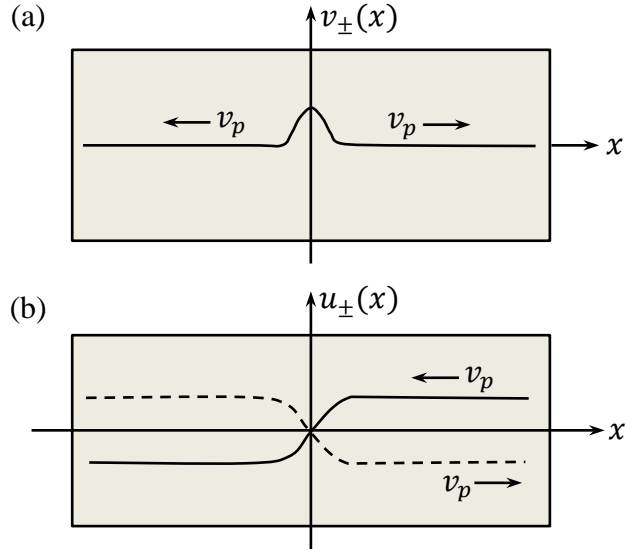

**Fig.2**. Plots of $v_\pm(x)$ and $u_\pm(x)$ at $t = 0$ inside an infinitely-wide slab. The plot of $u_-(x)$, shown with broken lines, is similar to that of $u_+(x)$, except for being flipped around the horizontal axis. As time progresses, $u_+(x)$ and $v_+(x)$ move to the left, while $u_-(x)$ and $v_-(x)$ move to the right, all at the constant velocity $v_p$. The overall displacement and velocity are $u(x,t) = u_+(x + v_p t) + u_-(x - v_p t)$ and $v(x,t) = v_+(x + v_p t) + v_-(x - v_p t)$, respectively.

As expected, we find $u(x, t = 0) = 0$, that is, the lattice has no dislocations at $t = 0$. However, recalling that $\omega = v_p k$, we will have

$$\partial_t u(x, t = 0) = \tfrac{1}{2} w_0 v_0 \int_{-\infty}^{\infty} \exp(-w_0|k|) \exp(ikx) \, dk = \frac{v_0}{1+(x/w_0)^2}. \tag{7}$$

In other words, the lattice has an initial velocity in a region of width $\sim \pi w_0$ centered at the origin of coordinates, as originally postulated. The linear momentum $p_x$ of the lattice along the $x$-axis is thus given by

$$p_x(t = 0) = \int_{-\infty}^{\infty} \frac{\mu v_0}{1+(x/w_0)^2} \, dx = \pi w_0 \mu v_0. \tag{8}$$

As time progresses, the initial momentum causes the following lattice deformation:

$$u(x, t) = \tfrac{1}{2}(w_0 v_0 / v_p) \int_{-\infty}^{\infty} k^{-1} \exp(-w_0|k|) \exp(ikx) \sin(k v_p t) \, dk$$

$$= \tfrac{1}{2}(w_0 v_0 / v_p) \int_{-\infty}^{\infty} \frac{\exp(-w_0|k|)}{2ik} \{\exp[ik(x + v_p t)] - \exp[ik(x - v_p t)]\} dk$$

$$= \tfrac{1}{2}(w_0 v_0 / v_p)\{\arctan[(x + v_p t)/w_0] - \arctan[(x - v_p t)/w_0]\}. \tag{9}$$

In other words, the deformation starts at $x = 0$, then spreads to the right and to the left with constant velocity $v_p$. The velocity profile of the lattice at time $t$ is now given by

$$v(x, t) = \partial_t u(x, t) = \tfrac{1}{2} w_0 v_0 \int_{-\infty}^{\infty} \exp(-w_0|k| + ikx) \cos(k v_p t) \, dk$$

$$= \tfrac{1}{4} w_0 v_0 \int_{-\infty}^{\infty} \exp(-w_0|k|) \{\exp[ik(x + v_p t)] + \exp[ik(x - v_p t)]\} dk$$

$$= \frac{\tfrac{1}{2} v_0}{1+[(x+v_p t)/w_0]^2} + \frac{\tfrac{1}{2} v_0}{1+[(x-v_p t)/w_0]^2}. \tag{10}$$

Clearly, the initial momentum of the lattice is preserved, although it is now divided between two regions at the leading and trailing edges of the expanding pulse. In between these two edges, the individual atoms/molecules of the lattice are displaced by $\tfrac{1}{2} \pi w_0 v_0 / v_p$ along the $x$-axis.

**2.1. Energy content of the elastic wave**. In the absence of initial displacement at $t = 0$, the entire energy of the system is initially contained in the kinetic energy of atoms/molecules constituting the medium, that is,

$$\mathcal{E}_K(t = 0) = \int_{-\infty}^{\infty} \tfrac{1}{2} \mu \left[\frac{v_0}{1+(x/w_0)^2}\right]^2 dx = \tfrac{1}{2} \mu v_0^2 w_0 \int_{-\infty}^{\infty} \frac{dx}{(1+x^2)^2} = \tfrac{1}{2} \mu v_0^2 w_0 \int_{-\pi/2}^{\pi/2} \frac{d\theta}{1+\tan^2 \theta}$$

$$= \tfrac{1}{2} \mu v_0^2 w_0 \int_{-\pi/2}^{\pi/2} \cos^2 \theta \, d\theta = \tfrac{1}{4} \pi \mu v_0^2 w_0. \tag{11}$$

At later times, when $t \gg 0$, the kinetic energy is contained in two well-separated pulses given by the two terms on the right-hand side of Eq.(10). This, however, accounts for only one-half of the initial energy of the pulse; the remaining half is carried by the potential energy of elastic springs that connect adjacent atoms/molecules, as follows:

$$\mathcal{E}_P(t \gg 0) = \tfrac{1}{2}\sigma \int_{-\infty}^{\infty}[\partial_x u(x,t)]^2 dx = \tfrac{1}{2}\sigma \int_{-\infty}^{\infty}\left(\frac{\tfrac{1}{2}(v_0/v_p)}{1+[(x+v_p t)/w_0]^2} - \frac{\tfrac{1}{2}(v_0/v_p)}{1+[(x-v_p t)/w_0]^2}\right)^2 dx$$

$$\cong \tfrac{1}{4}\sigma(v_0^2/v_p^2)w_0 \int_{-\infty}^{\infty}\frac{dx}{(1+x^2)^2} = \tfrac{1}{8}\pi\mu v_0^2 w_0. \tag{12}$$

It must be obvious from the preceding analysis that, when the leading and trailing edges of the (expanding) pulse are less than fully separated, the kinetic energy would contain a cross-term, namely, the product of the two terms in Eq.(10), which cancels out the cross-term associated with the potential energy contained in the first line of Eq.(12).

**Example 2**. Consider a slab of homogeneous material having width $2L$ along the $x$-axis, and infinite dimensions along the $y$ and $z$ axes. The slab boundaries at $x = \pm L$ are fixed, so that $u(x = \pm L, t) = 0$ at all times $t$. Let $u(x, t = 0) = 0$ and $v(x, t = 0) = v_0/[1 + (x/w_0)^2]$ in the interval $-L \leq x \leq L$, the implicit assumption here being that $w_0 \ll L$.

In order to satisfy the boundary conditions at $x = \pm L$, the initial distributions $u_\pm(x, t = 0)$ and $v_\pm(x, t = 0)$ must be expanded in a Fourier series (rather than a Fourier integral). In the present example, where the walls of the slab at $x = \pm L$ are rigidly affixed to external mounts, the initial velocity profiles $v_\pm(x, t = 0)$ must have the periodic structure depicted in Fig.3(a). As time progresses, $v_+(x)$ moves to the left and $v_-(x)$ moves to the right, both at the constant velocity $v_p$. A quick glance at Fig.3(a) confirms that the walls at $x = \pm L$ remain motionless at all times.

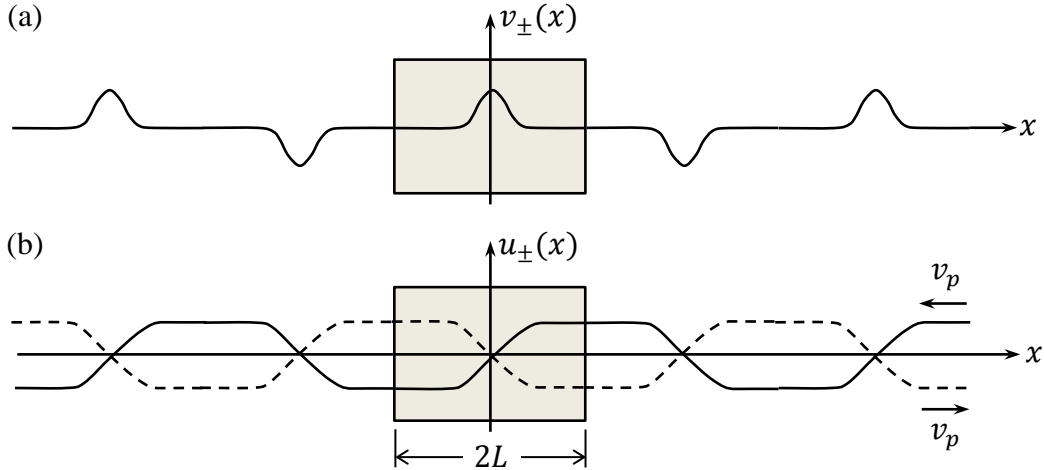

**Fig.3**. Plots of $v_\pm(x)$ and $u_\pm(x)$ at $t = 0$, when the slab's walls at $x = \pm L$ are rigidly affixed to external mounts. The plot of $u_-(x)$, shown with broken lines, is that of $u_+(x)$ flipped around the horizontal axis. As time progresses, $u_+(x)$ and $v_+(x)$ travel to the left, while $u_-(x)$ and $v_-(x)$ move to the right, all at the constant velocity $v_p$. Observe that at the boundaries $x = \pm L$, the overall displacement $u(x, t) = u_+(x + v_p t) + u_-(x - v_p t)$ and velocity $v(x, t) = v_+(x + v_p t) + v_-(x - v_p t)$ remain zero at all times.

In the absence of dispersion, one may obtain the displacement function $u_+(x, t = 0)$ by integrating the velocity profile $v_+(x, t = 0)$ over $x$; the result is shown in Fig.3(b). Clearly, the derivative with respect to $x$ of $u_+(x, t = 0)$ is equal to $v_+(x, t = 0)$. Considering that the initial displacement is zero, we will have $u_-(x, t = 0) = -u_+(x, t = 0)$. Now, the periodic function $v(x, t = 0)$, may be written

$$v(x, t = 0) = \left[\frac{v_0}{1+(x/w_0)^2}\right] * \left[\frac{1}{2L}\text{comb}\left(\frac{x}{2L}\right)\cos\left(\frac{\pi x}{2L}\right)\right]. \quad (13)$$

With the aid of the convolution and multiplication theorems of the Fourier transform theory (see Appendix A), the Fourier series representation of $v(x, t = 0)$ is found to be

$$V_0(k) = \pi w_0 v_0 \exp(-w_0|k|)\left\{\text{comb}\left(\frac{Lk}{\pi}\right) * \left[\frac{1}{2}\delta\left(k + \frac{\pi}{2L}\right) + \frac{1}{2}\delta\left(k - \frac{\pi}{2L}\right)\right]\right\}$$

$$= (\pi^2 w_0 v_0/L) \exp(-w_0|k|) \sum_{n=-\infty}^{\infty} \delta[k - (n + \frac{1}{2})\pi/L]. \quad (14)$$

Consequently, the Fourier transform of $u_\pm(x, t = 0)$ will be given by

$$U_\pm(k) = \pm\frac{V_0(k)}{2ikv_p} = \pm\frac{\pi^2 w_0 v_0 \exp(-w_0|k|)}{2ikLv_p}\sum_{n=-\infty}^{\infty} \delta[k - (n + \frac{1}{2})\pi/L]. \quad (15)$$

The spatio-temporal evolution of $u(x, t)$ and $v(x, t)$ may now be determined as was done in the case of unbounded media. Of course, for a slab having a finite-width, the spatial and temporal frequencies $k_n = (n + \frac{1}{2})\pi/L$ and $\omega_n$ are discrete, but they continue to satisfy $\omega_n = v_p k_n$, where $v_p = \sqrt{\sigma/\mu}$ is the phase velocity of propagation inside the material medium.

$$u(x, t) = u_+(x, t) + u_-(x, t)$$

$$= \frac{1}{2\pi}\int_{-\infty}^{\infty} \frac{\pi^2 w_0 v_0 \exp(-w_0|k|)}{kLv_p}\left\{\sum_{n=-\infty}^{\infty} \delta[k - (n + \frac{1}{2})\pi/L]\right\}\exp(ikx)\sin(\omega t)\,dk$$

$$= \frac{\pi w_0 v_0}{Lv_p}\sum_{n=0}^{\infty}\frac{\sin(k_n v_p t)}{k_n}\exp(-w_0 k_n)\cos(k_n x); \qquad [k_n = (n + \frac{1}{2})\pi/L]. \quad (16)$$

Note that the elastic wave bouncing back and forth within the (finite-width) slab maintains its energy, even though its momentum reverses direction each time the lattice exchanges momentum with the rigid support mounts located at $x = \pm L$.

**Example 3**. Here we examine the finite-width slab of the preceding example in the absence of the constraints on its sidewalls at $x = \pm L$. In other words, the sidewalls are now completely free to move along the $x$-axis. The corresponding plots of $v_\pm(x, t = 0)$ and $u_\pm(x, t = 0)$ are shown in Fig.4. The lattice retains its initial energy and momentum while moving forward along the $x$-axis, albeit in discrete steps. The initial velocity $v(x, t = 0)$, a periodic function, is given by

$$v(x, t = 0) = \left[\frac{v_0}{1+(x/w_0)^2}\right] * \left[\frac{1}{2L}\text{comb}\left(\frac{x}{2L}\right)\right]. \quad (17)$$

The Fourier transform of the above function is obtained with the aid of the convolution theorem (see Appendix A), as follows:

$$V_0(k) = \pi w_0 v_0 \exp(-w_0|k|)\text{comb}\left(\frac{Lk}{\pi}\right)$$

$$= (\pi^2 w_0 v_0/L)\exp(-w_0|k|)\sum_{n=-\infty}^{\infty}\delta(k - n\pi/L). \quad (18)$$

Consequently, the Fourier transform of $u_\pm(x, t = 0)$ will be given by

$$U_\pm(k) = \pm\frac{V_0(k)}{2ikv_p} = \pm\frac{\pi^2 w_0 v_0 \exp(-w_0|k|)}{2ikLv_p}\sum_{n=-\infty}^{\infty}\delta(k - n\pi/L). \quad (19)$$

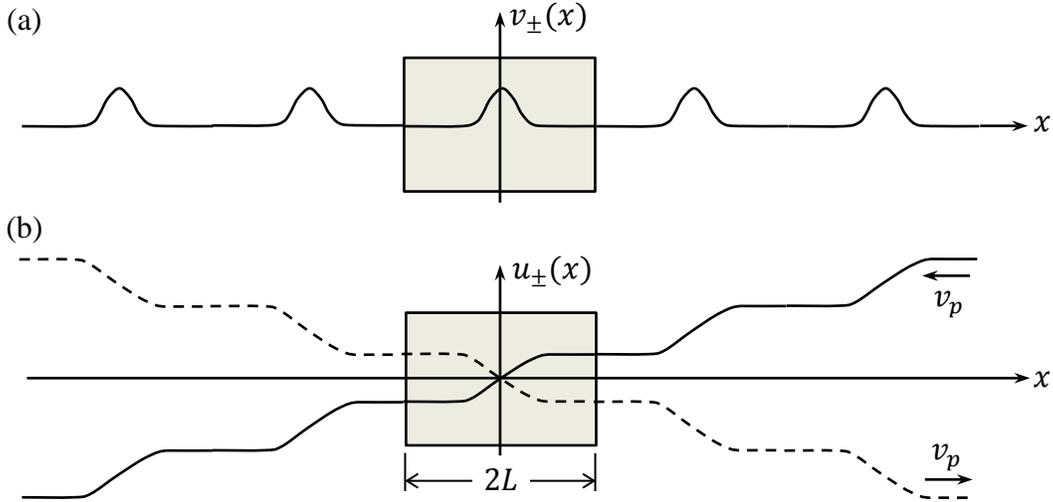

**Fig.4**. Plots of $v_\pm(x)$ and $u_\pm(x)$ at $t = 0$, when the slab's walls at $x = \pm L$ are unconstrained. The plot of $u_-(x)$, shown with broken lines, is similar to that of $u_+(x)$, except for being flipped around the horizontal axis. As time progresses, $u_+(x)$ and $v_+(x)$ move to the left, while $u_-(x)$ and $v_-(x)$ move to the right, all at the constant velocity $v_p$. Note that at the boundaries $x = \pm L$, the overall displacement $u(x,t) = u_+(x + v_p t) + u_-(x - v_p t)$ and velocity $v(x,t) = v_+(x + v_p t) + v_-(x - v_p t)$ do not vanish.

The spatio-temporal evolution of $u(x,t)$ and $v(x,t)$ may now be determined as was done in the case of unbounded media. Needless to say, the spatial and temporal frequencies $k_n = n\pi/L$ and $\omega_n = v_p k_n$ are discrete.

$$u(x,t) = u_+(x,t) + u_-(x,t)$$

$$= \frac{1}{2\pi} \int_{-\infty}^{\infty} \frac{\pi^2 w_0 v_0 \exp(-w_0|k|)}{kLv_p} \sum_{n=-\infty}^{\infty} \delta(k - n\pi/L) \exp(ikx) \sin(\omega t)\, dk$$

$$= \left(\frac{\pi w_0 v_0}{2L}\right) t + \frac{\pi w_0 v_0}{L v_p} \sum_{n=1}^{\infty} \frac{\sin(k_n v_p t)}{k_n} \exp(-w_0 k_n) \cos(k_n x); \quad (k_n = n\pi/L). \quad (20)$$

The elastic wave bouncing back and forth inside the (finite-width) slab maintains not only its energy but also its linear momentum in the present example, as there are no rigid mounts at $x = \pm L$ with which the excited vibrational modes could exchange momentum.

**Example 4**. Figure 5 shows the case of an asymmetrically excited finite-width slab, having one fixed and one free boundary, and an initial velocity profile $v(x, t = 0)$. The periodic functions $v_\pm(x)$ and $u_\pm(x)$, constructed to satisfy the boundary conditions and to allow reflections at the boundaries, have a period of $8L$. As expected, the material displacement at each point within the slab changes sign repeatedly and periodically, while the total energy of the system is conserved.

**Example 5**. Figure 6 provides another example of an asymmetrically excited finite-width slab, having an initial velocity profile $v(x, t = 0)$. Unlike the preceding example, the boundaries of the slab in the present case are free. The periodic (or quasi-periodic) functions $v_\pm(x)$ and $u_\pm(x)$, all having the same period, $4L$, are now constructed to satisfy the boundary conditions by allowing reflections without a sign change at the floating boundaries. The material displacements at each location within the slab are such that the slab in its entirety moves forward in a step-by-step fashion, while maintaining its total energy and linear momentum.

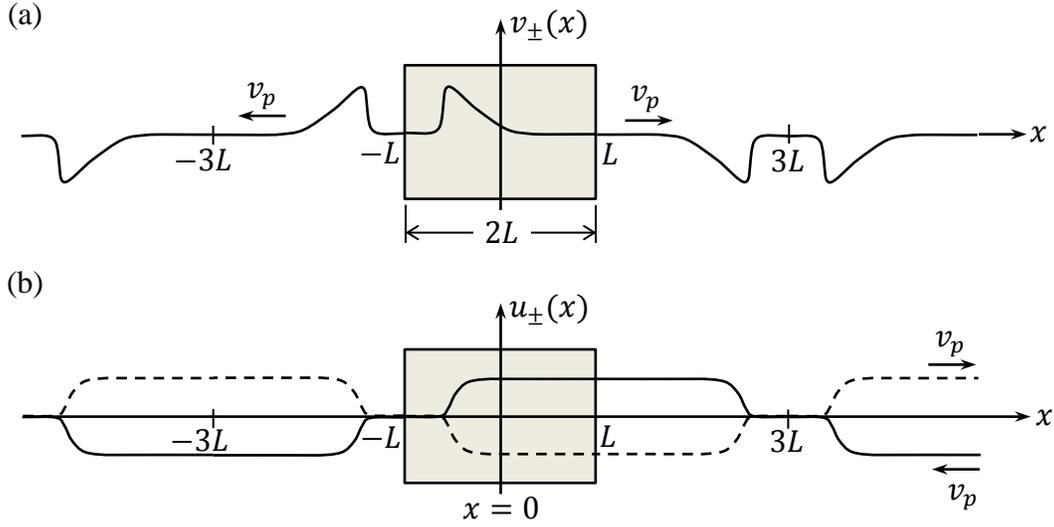

**Fig.5**. Plots of $v_\pm(x)$ and $u_\pm(x)$ at $t = 0$, when the initial disturbance is asymmetric, the slab's left wall at $x = -L$ is unconstrained, and the slab's right wall at $x = L$ is rigidly affixed to an external mount. The plot of $u_-(x)$, shown with broken lines, is obtained by flipping $u_+(x)$ around the $x$-axis. As time progresses, $u_+(x)$ and $v_+(x)$ move to the left, while $u_-(x)$ and $v_-(x)$ move to the right, all at the constant velocity $v_p$. Note that the overall displacement $u(x,t) = u_+(x + v_p t) + u_-(x - v_p t)$ and velocity $v(x,t) = v_+(x + v_p t) + v_-(x - v_p t)$ always vanish at the right-hand boundary. The free boundary on the left, however, oscillates back and forth at regular intervals.

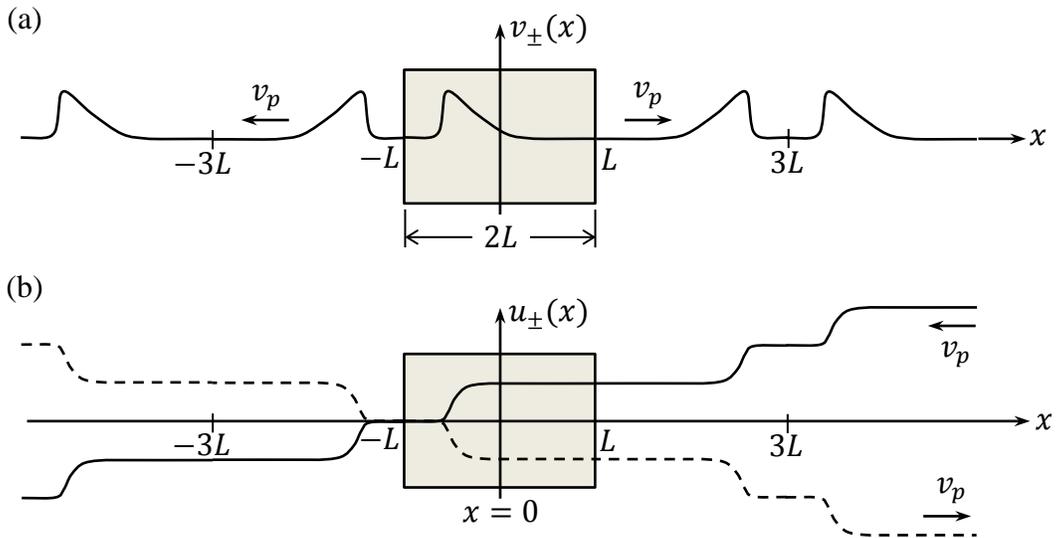

**Fig.6**. Plots of $v_\pm(x)$ and $u_\pm(x)$ at $t = 0$, when the initial disturbance is asymmetric and the slab's walls at $x = \pm L$ are unconstrained. The plot of $u_-(x)$, shown with broken lines, is obtained by flipping $u_+(x)$ around the $x$-axis. As time progresses, $u_+(x)$ and $v_+(x)$ move to the left, while $u_-(x)$ and $v_-(x)$ move to the right, all at the constant velocity $v_p$. Note that at the boundaries $x = \pm L$, the overall displacement $u(x,t) = u_+(x + v_p t) + u_-(x - v_p t)$ and velocity $v(x,t) = v_+(x + v_p t) + v_-(x - v_p t)$ do not vanish.

**2.2. Dispersive media.** As discussed in Sec.2.1, the general solution for elastic wave propagation in a uniform, homogeneous, infinite medium may be expressed as follows:

$$u(x,t) = \frac{1}{2\pi}\int_{-\infty}^{\infty}\{U_+(k)\exp[i(kx+\omega t)] + U_-(k)\exp[i(kx-\omega t)]\}dk, \quad (21a)$$

$$v(x,t) = \frac{1}{2\pi}\int_{-\infty}^{\infty}\{i\omega U_+(k)\exp[i(kx+\omega t)] - i\omega U_-(k)\exp[i(kx-\omega t)]\}dk. \quad (21b)$$

Fourier transforming the initial conditions $u(x,t=0)$ and $v(x,t=0)$ into $U_0(k)$ and $V_0(k)$, the functions $U_+(k)$ and $U_-(k)$ can be determined as follows:

$$U_+(k) = \tfrac{1}{2}[U_0(k) - iV_0(k)/\omega], \quad (22a)$$

$$U_-(k) = \tfrac{1}{2}[U_0(k) + iV_0(k)/\omega]. \quad (22b)$$

In order to introduce dispersion, one would like to relate $\omega$ to $k$ via some arbitrary function, say, $\omega(k)$, then continue to use Eqs.(21) and (22) to determine the temporal evolution of the initial conditions. This particular generalization of the solution of the wave equation, which now includes dispersion, does not pose any problems as far as momentum conservation is concerned; the reason being that the linear momentum of the medium is given by $p_x(t) = \mu\int_{-\infty}^{\infty} v(x,t)dx$, which, in accordance with the central value theorem (see Appendix A), is equal to $\mu$ times the Fourier transform of $v(x,t)$ at $k=0$. Invoking Eq.(21b), we write

$$p_x(t) = i\omega(0)\mu U_+(0)\exp[i\omega(0)t] - i\omega(0)\mu U_-(0)\exp[-i\omega(0)t]$$

$$= \mu V_0(0)\cos[\omega(0)t] - \mu\omega(0)U_0(0)\sin[\omega(0)t]. \quad (23)$$

The momentum of the system will, therefore, be time-independent if $\omega(0)$ happens to be zero, which is generally the case. We conclude that dispersion does not violate the conservation of momentum. Next, we examine the kinetic energy of the system $\mathcal{E}_K(t) = \tfrac{1}{2}\mu\int_{-\infty}^{\infty} v^2(x,t)dx$. Using Parseval's theorem (see Appendix A) in conjunction with Eq.(21b), we find

$$\mathcal{E}_K(t) = \frac{\mu}{4\pi}\int_{-\infty}^{\infty}|i\omega U_+(k)\exp(i\omega t) - i\omega U_-(k)\exp(-i\omega t)|^2 dk$$

$$= \frac{\mu}{4\pi}\int_{-\infty}^{\infty}|V_0(k)\cos(\omega t) - \omega(k)U_0(k)\sin(\omega t)|^2 dk. \quad (24)$$

As for the potential energy, we define $\mathcal{E}_P(t) = \tfrac{1}{2}\int_{-\infty}^{\infty}[\sigma(x)*u(x,t)]^2 dx$, where $*$ signifies the convolution operation, and $\sigma(x)$ is a real-valued weight function to be determined shortly. Invoking Parseval's theorem and denoting the Fourier transform of $\sigma(x)$ by $\Sigma(k)$, we write

$$\mathcal{E}_P(t) = \frac{1}{4\pi}\int_{-\infty}^{\infty}|\Sigma(k)U_+(k)\exp(i\omega t) + \Sigma(k)U_-(k)\exp(-i\omega t)|^2$$

$$= \frac{1}{4\pi}\int_{-\infty}^{\infty}|[\Sigma(k)/\omega(k)]V_0(k)\sin(\omega t) + \Sigma(k)U_0(k)\cos(\omega t)|^2. \quad (25)$$

Comparison of Eqs.(24) and (25) reveals that the total energy $\mathcal{E}_K + \mathcal{E}_P$ will be time-independent provided that

$$\Sigma(k) = i\sqrt{\mu}\,\omega(k). \quad (26)$$

Since the mass-density $\mu$ and the oscillation frequency $\omega(k)$ are real-valued, we conclude that $\sigma(x)$ must be an odd function of $x$, which in turn requires that $\omega(k)$ be an odd function of $k$. In the special case of a dispersionless medium, $\omega(k) = v_p k$ yields $\sigma(x) = \sqrt{\mu} v_p \delta'(x)$. We have

$$\mathcal{E}_P(t) = \tfrac{1}{2} \int_{-\infty}^{\infty} [\sigma(x) * u(x,t)]^2 dx = \tfrac{1}{2} \mu v_p^2 \int_{-\infty}^{\infty} [\partial_x u(x,t)]^2 dx. \tag{27}$$

Thus we have recovered the result for the simple case studied in Sec.1, where $\mu v_p^2$ is the spring constant $\sigma$ of the simple (i.e., dispersionless) material.

### 3. Transverse vibrations of a one-dimensional slab of homogeneous and uniform material.

A one-dimensional string stretched along the $x$-axis and vibrating along the transverse direction $y$, provides a good model for the transverse vibrations of a homogeneous, uniform slab of material. The diagram in Fig.7 helps to explain that the equation of motion along the $y$-axis is identical with that for longitudinal vibrations given by Eq.(1), provided that the tensile stress $\tau$ within the string is substituted for the spring constant $\sigma$ in Eq.(1). This is because, assuming the vibration amplitude is not too great, the slope of the displacement curve $\partial_x u(x,t) = \tan\theta$ may be approximated with $\sin\theta$. The vertical component of the tensile force acting within the wire at location $x$ and time $t$ is then written as $f_y(x,t) = \pm\tau \sin\theta \cong \pm\tau \partial_x u(x,t)$. The tensile stress $\tau$ thus plays the same role in transverse vibrations of a string as the stiffness coefficient $\sigma$ does in longitudinal vibrations. The phase velocity of propagation along the string is then $v_p = \sqrt{\tau/\mu}$.

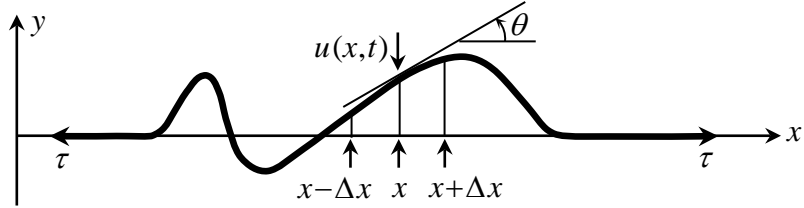

**Fig.7**. Transverse vibrations of a one-dimensional wire of uniform mass-density $\mu$ subject to a constant tension $\tau$. The amplitude of the vibrations is small enough to allow the approximation $\sin\theta \cong \partial_x u(x,t)$.

### 3.1. Potential energy of the string.

Suppose a short segment $\Delta x$ of the string is stretched and expanded along the $y$-axis, so that its length has become $\sqrt{(\Delta x)^2 + (\Delta y)^2} \cong \Delta x + \tfrac{1}{2}(\Delta y)^2/\Delta x$. Considering that the elongation by $\Delta\ell \cong \tfrac{1}{2}(\Delta y)^2/\Delta x = \tfrac{1}{2}[\partial_x u(x,t)]^2 \Delta x$ takes place under the constant tension $\tau$, the potential energy per unit-length stored in the string must be given by $\mathcal{E}_P(x,t) = \tfrac{1}{2}\tau[\partial_x u(x,t)]^2$. Once again, it is seen that the tensile stress $\tau$ plays the same role in transverse vibrations of the string as the spring constant $\sigma$ does in longitudinal vibrations.

### 3.2. Similarities and differences between longitudinal and transverse vibrations.

For an infinitely long string aligned with the $x$-axis and under a constant tensile stress $\tau$, transverse vibrations in the $y$ direction behave similarly to longitudinal vibrations along $x$. (A short pulse of light passing through can deposit its momentum and set off the vibrations.) All the $k$-vectors will be along the $x$-axis, but of course the overall momentum of the string will be directed along $y$.

Also, for a finite-length string with *fixed* end-points, the transverse behavior is similar to the longitudinal behavior. The energy of the string remains conserved, but its momentum continually gets exchanged with the fixed support structures to which the end-points are attached.

In the case of a finite-length string with *free* boundaries, one must create a contraption, such as that shown in Fig.8, in order to maintain a tensile stress within the string as the string moves away from its initial location. Also, if the initial velocity profile happens to be *asymmetric* relative to the center of the string, conservation of angular momentum will add a spinning motion to an (untethered) string, which would complicate the analysis. The contraption of Fig.8 allows periodic exchanges of angular momentum between the frame and the string, in such a way as to maintain the total angular momentum of the system while preventing the spinning of the string.

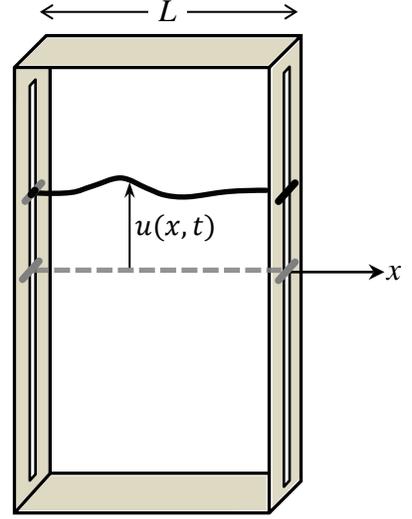

**Fig.8**. A string of length $L$ and negligible diameter is kept under a constant tensile stress $\tau$ by being mounted, tightly and horizontally, between the side-walls of a rectangular frame. The sliding mounts on both ends of the string are frictionless, leaving the string free to move up or down as the need arises. An initial upward kick, described by the initial conditions $u(x, t = 0)$ and $v(x, t = 0)$, is imparted to the string, starting it on an upward journey. The mounts, being essentially massless and frictionless, maintain the tensile stress $\tau$ along the length of the string at all times, while allowing the end-points to behave as if they were free to move in the vertical direction. No torque acts on the frame whenever the end-points of the string happen to be at the same height. However, at those instants when one end of the string rises above the other, the tensile force within the string exerts a torque that tends to rotate the frame — albeit ever so slightly, because of the frame's large moment of inertia. The string thus maintains its energy and linear momentum, while the system as a whole maintains its angular momentum by frequent exchanges between the string and the frame.

In accordance with Eq.(9), when a localized displacement of the string of Fig.8 reaches a free end-point and reverses direction, the end-point rises vertically by $\Delta y = \pi w_0 v_0 / v_p$. (A factor of 2 has been incorporated into this result because the height at the end-point doubles upon a reversal in the wave's propagation direction.) If one end of the string rises before the other end, the torque acting on the frame will be $\tau \Delta y = \pi w_0 \mu v_0 v_p$. This torque precisely accounts for the time rate-of-change of angular momentum of the vibrational motion of the string during the time when the leading and trailing edges of a deformation travel in the same direction — see Eq.(8), which gives the momentum of the entire spring along the $y$-axis, namely, $p_y = \pi w_0 \mu v_0$. When the leading and trailing edges travel in the same direction with velocity $v_p$, the time rate-of-change of angular momentum, $p_y v_p$, is precisely cancelled out by the torque acting on the frame.

**4. Solid object experiencing a constant external force**. Returning to longitudinal motion and one-dimensional excitation along the $x$-axis, let a constant, uniform force $F_0 \hat{x}$ be applied on the left-hand side of a solid rod of length $L$, linear mass-density $\mu$, and stiffness coefficient $\sigma$, as depicted in Fig.9. The situation is akin to a front-surface mirror subject to radiation pressure from a continuous wave (cw) light beam impinging from the left-hand-side. For a perfectly electrically conducting mirror, the light gets fully reflected at the front facet, which is the only place on which the external force acts. In the steady state, the rod will be slightly compressed — in order to produce the necessary internal forces along its length — and moves as a whole with constant acceleration $\boldsymbol{a} = (F_0/\mu L)\hat{x}$. The displacement function may thus be written as follows:

$$u(x,t) = \frac{F_0}{2L}\left[\frac{(x-L)^2}{\sigma} + \frac{t^2}{\mu}\right]; \qquad 0 \leq x \leq L, \ t \gg 0. \tag{28}$$

Note that the above $u(x,t)$ satisfies the wave equation $\mu \partial_t^2 u(x,t) = \sigma \partial_x^2 u(x,t)$ throughout the rod at all times $t$. Moreover, the rod's internal stress $\sigma \partial_x u(x,t) = (F_0/L)(x-L)$ vanishes at $x = L$, but equals $-F_0$ at $x = 0$, as it must.

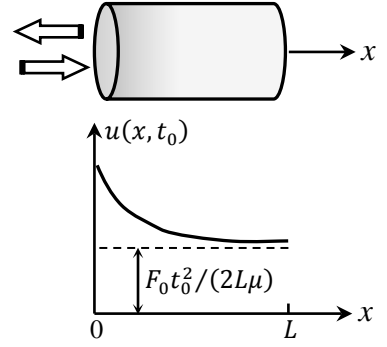

**Fig.9**. Uniform rod of length $L$, mass-density $\mu$, and stiffness coefficient $\sigma$ subject to a constant external force $F_0 \hat{x}$ on its left-hand facet. In the steady state, when all transient internal vibrations have settled down, the rod will be uniformly accelerating while slightly compressed, as dictated by the quadratic displacement function $u(x,t)$ depicted here.

Next, let us examine the case of an external force-density $\boldsymbol{f}(x) = f_0 \sin(2\pi x/L)\hat{x}$ acting within the body of a rod of length $L$. Since the total force $\boldsymbol{F} = \int_0^L \boldsymbol{f}(x)dx$ is zero in this case, the rod will not move as a whole, but it will develop internal stresses to cancel out the external force. This is a model for a transparent dielectric slab of refractive index $n$ and thickness $L$, illuminated at normal incidence by a linearly-polarized, monochromatic plane-wave of amplitude $E_0 \hat{y}$ and wavelength $\lambda_0$; see Appendix B for details. If the slab thickness happens to be $L = \lambda_0/2n$, the EM force distribution within the slab will be

$$\boldsymbol{f}(x) = -\frac{\pi \varepsilon_0 E_0^2 (n^2-1)^2}{2n\lambda_0} \sin\left(\frac{4\pi nx}{\lambda_0}\right)\hat{x}. \tag{29}$$

In the above equation, $\varepsilon_0$ is the permittivity of free space. The displacement function is now obtained as the second integral of the external force-density, namely,

$$u(x,t) = \frac{f_0 L}{2\pi\sigma}\left[\frac{L}{2\pi}\sin\left(\frac{2\pi x}{L}\right) - x\right]. \tag{30}$$

Note that the integration constants have been chosen to ensure that the internal stress $\sigma \partial_x u(x,t)$ vanishes at both boundaries located at $x = 0$ and $x = L$. The plot in Fig.10 shows the displacement function of Eq.(30) for the case of a transparent dielectric slab under cw illumination, where $f_0 < 0$; see Eq.(29).

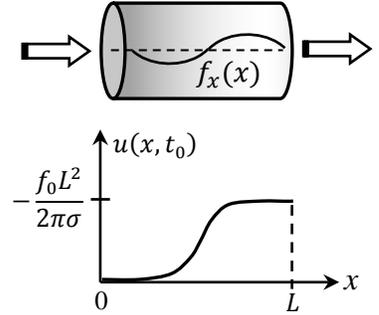

**Fig.10**. A dielectric slab of refractive index $n$ and thickness $L = \lambda_0/2n$ is illuminated from the left-hand-side by a plane-wave of wavelength $\lambda_0$. The radiation force acting on the dielectric medium is given by Eq.(29), and the resulting material displacement is given by Eq.(30). Since the integrated force of radiation on the slab is zero, the function $u(x,t)$ is time-independent. Note that the curvature of the left half of the displacement function is positive, while that of the right half is negative. The internal stresses of the slab thus precisely cancel out the external force exerted by the EM field.

Our next example pertains once again to a solid, uniform rod of length $L$, but now subject to the external force-density $\boldsymbol{f}(x) = f_0 \sin(\pi x/L)\hat{x}$. The total force $\boldsymbol{F} = \int_0^L \boldsymbol{f}(x)dx = (2f_0 L/\pi)\hat{x}$ being nonzero in the present example, the rod as a whole will acquire an acceleration $\boldsymbol{a} = 2f_0\hat{x}/(\pi\mu)$, but it will also develop internal stresses to cancel out the residual external force. This is a model for a transparent dielectric slab of refractive index $n$ and thickness $L = \lambda_0/4n$, illuminated at normal incidence by a linearly polarized, monochromatic plane-wave of amplitude $E_0\hat{y}$ and wavelength $\lambda_0$, for which the EM force-density, as shown in Appendix B, is given by

$$f(x) = \frac{2\pi n \varepsilon_0 E_0^2}{\lambda_0} \left(\frac{n^2-1}{n^2+1}\right)^2 \sin\left(\frac{4\pi n x}{\lambda_0}\right) \hat{x}. \tag{31}$$

The displacement function is now obtained as the second integral of the residual external force-density, that is,

$$u(x,t) = \frac{f_0 L}{\pi \sigma}\left[\frac{L}{\pi}\sin\left(\frac{\pi x}{L}\right) + \frac{x}{L}(x-L)\right] + \left(\frac{f_0}{\pi \mu}\right) t^2, \tag{32}$$

where the integration constants have been chosen to ensure that the internal stress $\sigma \partial_x u(x,t)$ vanishes at both ends of the rod located at $x = 0$ and $x = L$, while the average force-density $(1/L)\int_0^L \sigma \partial_x^2 u(x,t) dx$ survives. The plot of $u(x,t)$ of Eq.(32) shown in Fig.11 corresponds to a quarter-wave-thick dielectric slab under cw illumination, for which $f_0 > 0$; see Eq.(31).

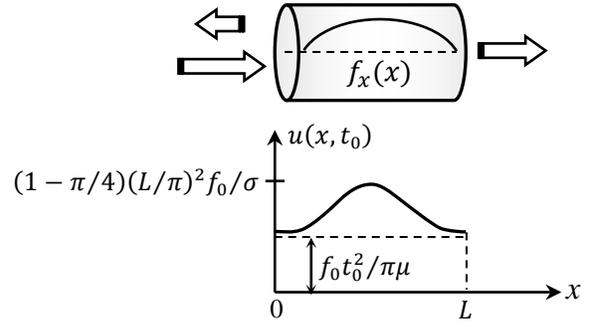

**Fig.11**. Dielectric slab of refractive index $n$ and thickness $L = \lambda_0/4n$, illuminated from the left-hand-side by a plane-wave of wavelength $\lambda_0$. The radiation force on the dielectric is given by Eq.(31), and the resulting material displacement is given by Eq.(32). The integrated force of radiation being positive, the slab moves forward with a net acceleration. Note that the displacement function's curvature is negative in the mid-section and positive near the entrance and exit facets of the slab. The internal stresses of the slab thus precisely cancel out the residual external force exerted by the EM field.

Finally, we consider the case of an antireflection coating layer atop a semi-infinite substrate of refractive index $n$. The layer has index of refraction $\sqrt{n}$ and thickness $L = \lambda_0/(4\sqrt{n})$, where $\lambda_0$ is the wavelength of the normally-incident plane-wave of amplitude $E_0 \hat{y}$. In accordance with the analysis in Appendix B, the EM force-density acting within the coating layer is

$$f(x) = -\frac{\pi (n-1)^2 \varepsilon_0 E_0^2}{2\sqrt{n}\lambda_0} \sin\left(\frac{4\pi \sqrt{n} x}{\lambda_0}\right) \hat{x}. \tag{33}$$

We thus consider the external force-density $f(x) = f_0 \sin(\pi x/L)\hat{x}$, acting inside a one-dimensional rod of length $L$, as shown in Fig.12. The second integral of this force-density yields the displacement function of the rod, as follows:

$$u(x,t) = \frac{f_0 L}{\pi \sigma}\left[\frac{L}{\pi}\sin(\pi x/L) + (L-x)\right]. \tag{34}$$

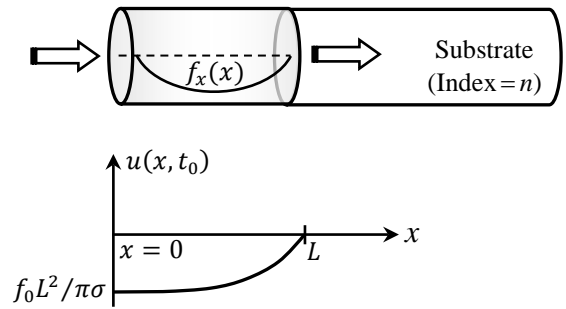

**Fig.12**. Antireflection coating layer of refractive index $\sqrt{n}$ and thickness $L = \lambda_0/(4\sqrt{n})$ ensures that an incident plane-wave of wavelength $\lambda_0$ enters the substrate with no reflection losses at the entrance facet. The EM force density inside the layer is given by Eq.(33), and the corresponding displacement function appears in Eq.(34). The internal force-density cancels out the EM force-density throughout the coating layer, while the slope of $u(x,t)$ at $x = L$ guarantees that the total EM force $F = (2f_0 L/\pi)\hat{x}$ pulls at the interface between the coating layer and the substrate.

Note that the integration constants have been chosen to ensure that the internal force completely cancels the external force, that the internal stress $\sigma \partial_x u(x,t)$ at $x = 0$ vanishes, that the internal stress at $x = L$ pulls at the substrate interface with a force of $2f_0 L/\pi$ (this is a pull force when $f_0 < 0$), and that the displacement at $x = L$ is equal to zero, since the substrate has been taken to be immobile.

**5. Equation of motion for one-dimensional rod under internal stress**. So far, we have examined fairly simple situations involving either longitudinal or transverse motion in one dimension associated with thin solid rods. In the present section we analyze more complex mechanical vibrations and deformations of a one-dimensional medium that is free to move in three dimensional space. We will find that, although the resulting equation of motion does *not* admit simple analytical solutions, it is nonetheless possible to verify that its solutions comply with the conservation laws of energy and linear as well as angular momentum.

Consider a thin, solid rod of length $L$, mass-density $\mu(\rho)$, and stiffness coefficient $\sigma(\rho)$, as shown in Fig.13. Here $\rho$ is the distance from the left-end of the rod when the rod is at rest. The rod is initially at rest on the $y$-axis, occupying the interval $[0, L]$. Denoting the position of the point $\rho$ of the rod at time $t$ with the vector function $\boldsymbol{u}(\rho, t)$, the equation of motion of the rod may be written as follows:

$$\partial_\rho \left[ \sigma(\rho) \left( \partial_\rho \boldsymbol{u} - \frac{\partial_\rho \boldsymbol{u}}{|\partial_\rho \boldsymbol{u}|} \right) \right] = \mu(\rho) \partial_t^2 \boldsymbol{u}(\rho, t). \tag{35}$$

The mass-density $\mu(\rho)$ and the stiffness coefficient $\sigma(\rho)$ of the rod are positive everywhere. Note that $\partial_\rho \boldsymbol{u}(\rho, t)$ represents both the normalized length and orientation at time $t$ of the infinitesimal length of the spring initially located at $\rho$. From this, the unit-vector $\partial_\rho \boldsymbol{u}/|\partial_\rho \boldsymbol{u}|$ has been subtracted in Eq.(35) to yield the normalized elongation of the local spring at time $t$. Multiplication by $\sigma(\rho)$ yields the tensile force within the spring, and the final differentiation with respect to $\rho$ results in the net force (per unit length) acting on the local mass-density $\mu(\rho)$.

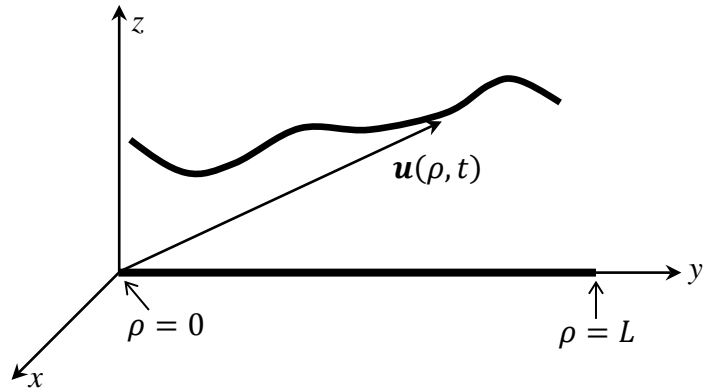

**Fig.13**. When $t < 0$, the one-dimensional rod of length $L$ and negligible thickness rests on the $y$-axis, overlaying the interval $[0, L]$. Each point on the rod is uniquely identified by its distance $\rho$ from the left-end of the rod in its rest state. The rod's mass-density and stiffness coefficient are denoted by $\mu(\rho)$ and $\sigma(\rho)$, respectively. At times $t \geq 0$, the location of the point $\rho$ on the rod in three-dimensional Euclidean space is specified by the vector $\boldsymbol{u}(\rho, t)$; the rod's instantaneous velocity profile is thus $\boldsymbol{v}(\rho, t) = \partial_t \boldsymbol{u}(\rho, t)$. The initial conditions are specified as $\boldsymbol{u}(\rho, t = 0)$ and $\boldsymbol{v}(\rho, t = 0)$.

The overall linear momentum of the rod at time $t$ may now be evaluated as follows:

$$\boldsymbol{p}(t) - \boldsymbol{p}(0) = \int_{t'=0}^{t} \int_{\rho=0}^{L} \mu(\rho) \partial_t^2 \boldsymbol{u}(\rho, t') d\rho dt'$$

$$= \int_{t'=0}^{t} \int_{\rho=0}^{L} \partial_\rho \left[ \sigma(\rho) \left( \partial_\rho \boldsymbol{u} - \frac{\partial_\rho \boldsymbol{u}}{|\partial_\rho \boldsymbol{u}|} \right) \right] d\rho dt'$$

$$= \int_{t'=0}^{t} \left[ \sigma(L) \left( \partial_\rho \boldsymbol{u} - \frac{\partial_\rho \boldsymbol{u}}{|\partial_\rho \boldsymbol{u}|} \right) \bigg|_{\rho=L} - \sigma(0) \left( \partial_\rho \boldsymbol{u} - \frac{\partial_\rho \boldsymbol{u}}{|\partial_\rho \boldsymbol{u}|} \right) \bigg|_{\rho=0} \right] dt'. \quad (36)$$

Conservation of linear momentum thus demands that $|\partial_\rho \boldsymbol{u}(0, t)| = |\partial_\rho \boldsymbol{u}(L, t)| = 1.0$ at all times. These are the boundary conditions for a rod whose end-points are free to move about.

As for the conservation of angular momentum for an unconstrained rod, it suffices to show that the overall torque acting on the rod remains zero at all times. We will have

$$\boldsymbol{T}(t) = \int_0^L \boldsymbol{u}(\rho, t) \times \partial_\rho \left[ \sigma(\rho) \left( \partial_\rho \boldsymbol{u} - \frac{\partial_\rho \boldsymbol{u}}{|\partial_\rho \boldsymbol{u}|} \right) \right] d\rho$$

$$\boxed{\text{Integration by parts}} \rightarrow = \boldsymbol{u}(L, t) \times \sigma(L) \left( \partial_\rho \boldsymbol{u} - \frac{\partial_\rho \boldsymbol{u}}{|\partial_\rho \boldsymbol{u}|} \right) \bigg|_{\rho=L} - \boldsymbol{u}(0, t) \times \sigma(0) \left( \partial_\rho \boldsymbol{u} - \frac{\partial_\rho \boldsymbol{u}}{|\partial_\rho \boldsymbol{u}|} \right) \bigg|_{\rho=0}$$

$$- \int_0^L \partial_\rho \boldsymbol{u}(\rho, t) \times \sigma(\rho) [1 - |\partial_\rho \boldsymbol{u}(\rho, t)|^{-1}] \partial_\rho \boldsymbol{u}(\rho, t) d\rho. \quad (37)$$

When the end-points of the rod are free to move, we will have $|\partial_\rho \boldsymbol{u}(0, t)| = |\partial_\rho \boldsymbol{u}(L, t)| = 1.0$, which causes the first two terms on the right-hand side of Eq.(37) to vanish. The remaining term also vanishes because the cross-product of $\partial_\rho \boldsymbol{u}(\rho, t)$ into itself is equal to zero. The conservation of angular momentum for a free-floating rod is thus confirmed.

Next, we consider the energy of the rod as a function of time, and demonstrate that the sum of its kinetic and potential energies remain constant regardless of whether the rod is free-floating, fixed at one end-point, or fixed at both end-points. The kinetic and potential energies of the rod at time $t$ are written as follows:

$$\mathcal{E}_K(t) = \tfrac{1}{2} \int_0^L \mu(\rho) v^2(\rho, t) d\rho = \tfrac{1}{2} \int_0^L \mu(\rho) \partial_t \boldsymbol{u}(\rho, t) \cdot \partial_t \boldsymbol{u}(\rho, t) d\rho. \quad (38)$$

$$\mathcal{E}_P(t) = \tfrac{1}{2} \int_0^L \sigma(\rho) \left( \partial_\rho \boldsymbol{u} - \frac{\partial_\rho \boldsymbol{u}}{|\partial_\rho \boldsymbol{u}|} \right) \cdot \left( \partial_\rho \boldsymbol{u} - \frac{\partial_\rho \boldsymbol{u}}{|\partial_\rho \boldsymbol{u}|} \right) d\rho. \quad (39)$$

Taking the time-derivative of the kinetic energy $\mathcal{E}_K(t)$ given by Eq.(38) and invoking the equation of motion, Eq.(35), we find

$$\tfrac{d}{dt} \mathcal{E}_K(t) = \int_0^L \mu(\rho) \partial_t^2 \boldsymbol{u}(\rho, t) \cdot \partial_t \boldsymbol{u}(\rho, t) d\rho$$

$$= \int_0^L \partial_\rho \left[ \sigma(\rho) \left( \partial_\rho \boldsymbol{u} - \frac{\partial_\rho \boldsymbol{u}}{|\partial_\rho \boldsymbol{u}|} \right) \right] \cdot \partial_t \boldsymbol{u}(\rho, t) d\rho$$

$$\boxed{\text{Integration by parts}} \rightarrow = \sigma(L)\left(\partial_\rho \boldsymbol{u} - \frac{\partial_\rho \boldsymbol{u}}{|\partial_\rho \boldsymbol{u}|}\right)\bigg|_{\rho=L} \cdot \partial_t \boldsymbol{u}(L,t) - \sigma(0)\left(\partial_\rho \boldsymbol{u} - \frac{\partial_\rho \boldsymbol{u}}{|\partial_\rho \boldsymbol{u}|}\right)\bigg|_{\rho=0} \cdot \partial_t \boldsymbol{u}(0,t)$$

$$- \int_0^L \sigma(\rho)\left(\partial_\rho \boldsymbol{u} - \frac{\partial_\rho \boldsymbol{u}}{|\partial_\rho \boldsymbol{u}|}\right) \cdot \frac{\partial^2 \boldsymbol{u}(\rho,t)}{\partial \rho \partial t} d\rho. \tag{40}$$

The first two terms on the right-hand-side of Eq.(40) vanish, whether the end-points of the rod are free or fully constrained (i.e., fixed). The remaining term is cancelled by the time-rate-of-change of the potential energy $\mathcal{E}_P(t)$, which is evaluated as follows:

$$\frac{d}{dt}\mathcal{E}_P(t) = \int_0^L \sigma(\rho)\left(\partial_\rho \boldsymbol{u} - \frac{\partial_\rho \boldsymbol{u}}{|\partial_\rho \boldsymbol{u}|}\right) \cdot \partial_t\left(\partial_\rho \boldsymbol{u} - \frac{\partial_\rho \boldsymbol{u}}{|\partial_\rho \boldsymbol{u}|}\right) d\rho$$

$$= \int_0^L \sigma(\rho)\left(\partial_\rho \boldsymbol{u} - \frac{\partial_\rho \boldsymbol{u}}{|\partial_\rho \boldsymbol{u}|}\right) \cdot \frac{\partial^2 \boldsymbol{u}(\rho,t)}{\partial \rho \partial t} d\rho. \tag{41}$$

The reason for removing the term $\partial_t(\partial_\rho \boldsymbol{u}/|\partial_\rho \boldsymbol{u}|)$ from Eq.(41) is that the unit-vector $\partial_\rho \boldsymbol{u}/|\partial_\rho \boldsymbol{u}|$ can only change in a direction perpendicular to itself, thus making $\partial_t(\partial_\rho \boldsymbol{u}/|\partial_\rho \boldsymbol{u}|)$ orthogonal to $(1 - |\partial_\rho \boldsymbol{u}|^{-1})\partial_\rho \boldsymbol{u}(\rho,t)$, causing the dot-product of the two vectors to vanish. We conclude that, irrespective of whether the end-points of the rod are free or fixed, its total energy is conserved.

**6. Elastic wave propagation in the transient regime**. Having observed the complications that could arise when elastic motion is unconstrained, we return to simpler problems involving longitudinal displacements and deformations in one dimension. This time, however, we change the focus of our attention to transient effects, which cannot be handled by simply introducing static deformations. The examples in the following subsections show how deformations initiate and then propagate through homogeneous semi-infinite rods in one-dimensional space.

**6.1. Semi-infinite rod under constant external pressure**. Shown in Fig.14 is a straight, uniform, homogeneous, semi-infinite rod aligned with the $x$-axis. The rod has cross-sectional area $A$, mass-density $\mu$, stiffness coefficient $\sigma$, and phase velocity $v_p = \sqrt{\sigma/\mu}$. Starting at $t = 0$, a uniform pressure $P_0$, acting on the left-hand facet of the rod, pushes the rod forward along the $x$-axis. The displacement of the rod $u(x,t)$ is given by the following function, which satisfies the wave equation $\sigma \partial_x^2 u(x,t) = \mu \partial_t^2 u(x,t)$, as it has the standard form of $u(x - v_p t)$.

$$u(x,t) = \begin{cases} -(v_0/v_p)(x - v_p t); & 0 \leq x \leq v_p t - \ell, \\ \alpha \sin^2[\beta(x - v_p t)]; & v_p t - \ell \leq x \leq v_p t, \\ 0; & x \geq v_p t. \end{cases} \tag{42}$$

In Eq.(42), $\ell$ is the width of the leading edge of the wavefront, which, together with the shape of the leading edge, is determined by the way in which the external pressure rises from zero to $P_0$ in the vicinity of $t = 0$. The functional form of the leading edge is chosen rather arbitrarily here, with the parameters $\alpha$ and $\beta$ intended to be adjusted to ensure a smooth transition from the linear segment of $u(x,t)$ to the nonlinear segment representing the leading edge. The velocity $v(x,t) = \partial_t u(x,t)$ is seen to be constant at $v = v_0$ in the linear segment between $x = 0$ and $x = v_p t - \ell$. The internal force acting on the left-hand facet of the rod, that is, $\sigma \partial_x u(x = 0, t) = -\sigma v_0 / v_p$, must be cancelled out by the external push force $F_x = AP_0$. Therefore, $v_0 = (AP_0/\sigma)v_p$ is the rod's velocity in the linear region $0 \leq x \leq v_p t - \ell$. At $x = v_p t - \ell$ both $u(x,t)$ and $\partial_x u(x,t)$ must be continuous, that is,

$$\begin{cases} \alpha \sin^2(\beta \ell) = \ell v_0 / v_p \\ 2\alpha\beta \sin(\beta\ell)\cos(\beta\ell) = v_0/v_p \end{cases} \rightarrow \begin{cases} \tan(\beta\ell) = 2\beta\ell \\ \alpha\beta \sin(2\beta\ell) = AP_0/\sigma \end{cases} \rightarrow \begin{cases} \beta\ell \cong 1.1655 \\ \alpha \cong 1.184(AP_0\ell/\sigma). \end{cases} \quad (43)$$

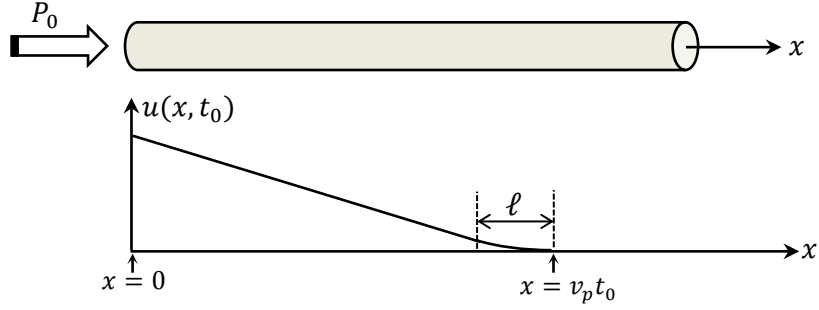

**Fig.14.** Starting at $t = 0$, a uniform, semi-infinite rod, having cross-sectional area $A$, mass-sensity $\mu$, and stiffness coefficient $\sigma$, is subjected to a constant pressure $P_0$ on its left-hand facet. In the absence of dispersion, the pressure wave inside the rod propagates at the constant velocity $v_p$, producing the displacement $u(x,t)$ along the length of the rod at times $t \geq 0$. The displacement is a linear function of $x$, except in the short interval $\ell$ immediately before the leading edge of the wavefront, where it has the nonlinear form given in Eq.(42). The nonlinearity is rooted in the fact that the pressure acting on the left-hand facet must rise from zero to $P_0$ in a short time interval in the vicinity of $t = 0$. At any given time, say, $t = t_0$, the displacement is largest at $x = 0$ and drops off linearly with increasing $x$ up to the point $x = v_p t_0 - \ell$, where it merges with the nonlinear section representing the leading edge of the wavefront, and proceeds to vanish smoothly at $x = v_p t_0$.

Thus, given the width $\ell$ of the leading edge, the parameters $\alpha$ and $\beta$ are readily obtained from Eq.(43). The linear momentum of the rod at time $t$ may now be calculated as follows:

$$p_x(t) = \int_0^\infty \mu v(x,t) dx = \int_0^{v_p t} \mu \partial_t u(x,t) dx$$

$$= \mu v_0(v_p t - \ell) + \mu\alpha\beta v_p \int_0^\ell \sin(2\beta x) dx = \mu v_0(v_p t - \ell) + \mu\alpha v_p \sin^2(\beta\ell)$$

$$= \mu v_0(v_p t - \ell) + \mu v_0 \ell = \mu v_0 v_p t = AP_0 t. \quad (44)$$

The above result may appear surprising at first, considering that the initial pressure during the period $0 \leq t \leq \ell/v_p$ has not been equal to $P_0$. Note, however, that the pressure history, which is inherent to the displacement profile of Eq.(42), is given by

$$P(t) = -(\sigma/A)\partial_x u(x = 0, t) = \begin{cases} 0; & t \leq 0, \\ (\alpha\beta\sigma/A)\sin(2\beta v_p t); & 0 \leq t \leq \ell/v_p, \\ \sigma v_0/Av_p; & t \geq \ell/v_p. \end{cases} \quad (45)$$

Considering that $2\beta\ell \cong 2.331 > \tfrac{1}{2}\pi$, it is clear that $P(t)$ rises above $P_0$ during the initial interval $0 \leq t \leq \ell/v_p$ before settling down at $P_0$. In general, any pressure function $P(t)$ produces its own displacement function $u(x - v_p t)$ via a simple integral, and the mechanical momentum of the rod can easily be shown to satisfy the following conservation law:

$$p_x(t) = \int_0^\infty \mu v(x,t) dx = \int_0^{v_p t} \mu \partial_t u(x - v_p t) dx = -\int_0^{v_p t} \mu v_p \partial_x u(x - v_p t) dx$$

$$= -\int_0^t \mu v_p^2 \partial_x u(v_p t'' - v_p t) dt'' = -\int_0^t \sigma \partial_x u(-v_p t') dt'$$

$$= -\int_0^t \sigma \partial_x u(x=0, t') dt' = A \int_0^t P(t') dt'. \tag{46}$$

To appreciate the generality of the above result, we give another example of the displacement function constructed from a simple $P(t)$, which rises linearly with time during the initial interval $0 \le t \le \ell/v_p$, before settling at the constant pressure $P_0$. We will have

$$u(x,t) = \begin{cases} \alpha - (v_0/v_p)(x - v_p t); & 0 \le x \le v_p t - \ell, \\ \beta(x - v_p t)^2; & v_p t - \ell \le x \le v_p t, \\ 0; & x \ge v_p t. \end{cases} \tag{47}$$

The continuity of $u(x,t)$ and $\partial_x u(x,t)$ at $x = v_p t - \ell$ now yields

$$\begin{cases} \alpha + (v_0/v_p)\ell = \beta \ell^2 \\ v_0/v_p = 2\beta\ell \end{cases} \rightarrow \begin{cases} \alpha = -\tfrac{1}{2}(v_0/v_p)\ell \\ \beta\ell = \tfrac{1}{2}(v_0/v_p). \end{cases} \tag{48}$$

The linear momentum of the rod at time $t$ is now easily calculated, as follows:

$$p_x(t) = \int_0^\infty \mu v(x,t) dx = \int_0^{v_p t} \mu \partial_t u(x,t) dx$$

$$= \mu v_0 (v_p t - \ell) + 2\mu v_p \beta \int_0^\ell x\, dx = \mu v_0 (v_p t - \ell) + \mu v_p \beta \ell^2$$

$$= \mu v_0 (v_p t - \ell) + \tfrac{1}{2} \mu v_0 \ell = A P_0 [t - \ell/(2 v_p)]. \tag{49}$$

Finally, we demonstrate that the energy of the rod is, in general, equally split between kinetic and potential energies. This is done by invoking the general formulas for the kinetic and potential energies, namely,

$$\mathcal{E}_K = \int_0^{v_p t} \tfrac{1}{2} \mu [\partial_t u(x - v_p t)]^2 dx = \int_0^{v_p t} \tfrac{1}{2} \mu v_p^2 [\partial_x u(x - v_p t)]^2 dx$$

$$= \int_0^{v_p t} \tfrac{1}{2} \sigma [\partial_x u(x - v_p t)]^2 dx = \mathcal{E}_P. \tag{50}$$

**6.2. Transparent semi-infinite medium illuminated by a light pulse**. Inside the transparent rod shown in Fig.15, the light beam's leading edge propagates at the speed of light $v_c$. The pressure of the light on the medium is confined to the vicinity of the leading edge, where the displacement function $u(x,t)$ has a non-zero curvature. Since no radiation pressure is exerted at the entrance facet of the medium, the slope of $u(x,t)$ at $x = 0$ must vanish. The general form of the displacement function shown in Fig.15 has the following mathematical expression:

$$u(x,t) = \begin{cases} 0; & x \geq v_c t, \\ f(x - v_c t); & v_c t - \ell_f \leq x \leq v_c t, \\ u_0 - (v_0/v_c)(x - v_c t); & v_p t \leq x \leq v_c t - \ell_f, \\ u_0 + (1 - v_p/v_c)v_0 t + g(x - v_p t); & v_p t - \ell_g \leq x \leq v_p t, \\ u_1 + (1 - v_p/v_c)v_0 t; & 0 \leq x \leq v_p t - \ell_g. \end{cases} \quad (51)$$

The continuity of $u(x,t)$ and its derivative $\partial_x u(x,t)$ requires that $f(0) = f'(0) = 0$, $f(-\ell_f) = u_0 + (v_0/v_c)\ell_f$, $f'(-\ell_f) = -(v_0/v_c)$, $g(0) = 0$, $g'(0) = -(v_0/v_c)$, $g(-\ell_g) = u_1 - u_0$, and $g'(-\ell_g) = 0$.

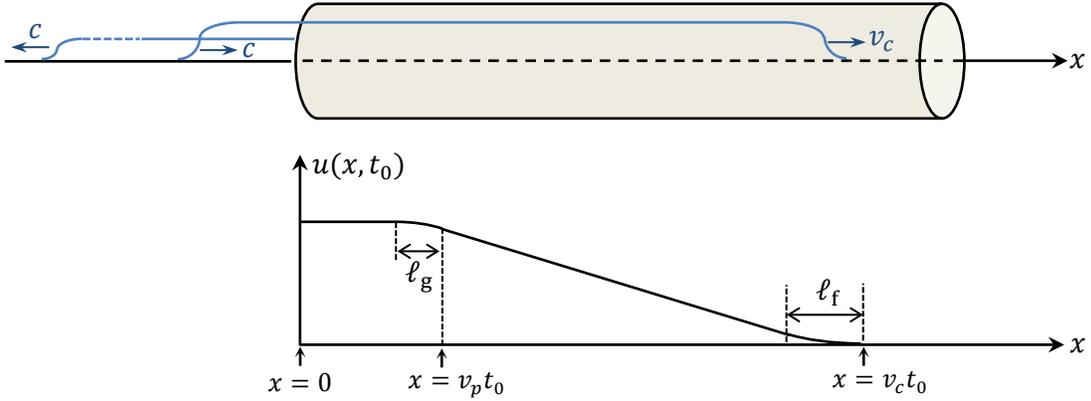

**Fig.15**. Light Pulse entering a semi-infinite, homogeneous, transparent, dispersionless rod from its left facet located at $x = 0$. Also shown is the fraction of the incident pulse reflected at the entrance facet. In the free space region outside the rod, the light propagates at velocity $c$, but inside the rod its velocity drops to $v_c = c/n$, where $n$ is the rod's refractive index. The leading-edge of the pulse exerts a force on the material medium, causing the one-dimensional motion described by the displacement function $u(x,t)$. While the intrinsic disturbances of the elastic medium propagate at the speed of sound, $v_p$, the mechanical motion induced by the leading-edge of the light pulse propagates at the much larger speed of light, $v_c$.

The function $u(x,t)$ of Eq.(51) satisfies the homogeneous wave equation everywhere except in the interval of length $\ell_f$ immediately before the leading edge of the light pulse, where the mass-density times the acceleration, $\mu \partial_t^2 u(x,t) = \mu v_c^2 f''(x - v_c t)$, is significantly greater than the internal force-density $\sigma \partial_x^2 u(x,t) = \mu v_p^2 f''(x - v_c t)$. The difference between these two expressions, namely, $\mu(v_c^2 - v_p^2)f''(x - v_c t)$, must of course be equal to the optical force-density exerted by the leading edge of the light beam. Integrating this optical force-density over the interval of length $\ell_f$ yields the total force exerted by the leading edge of the pulse as

$$F_x = \mu(v_c^2 - v_p^2)[f'(0) - f'(-\ell_f)] = \mu[1 - (v_p/v_c)^2]v_0 v_c. \quad (52)$$

Next, we compute the total momentum of the slab at time $t$, which is readily obtained from Eq.(51), as follows:

$p_x(t) = \int_0^\infty \mu \partial_t u(x,t) dx$

$$= \mu(1 - v_p/v_c)v_0(v_p t - \ell_g) + \int_{v_p t - \ell_g}^{v_p t} \mu[(1 - v_p/v_c)v_0 - v_p g'(x - v_p t)]dx$$

$$+ \mu v_0[(v_c - v_p)t - \ell_f] - \mu v_c \int_{v_c t - \ell_f}^{v_c t} f'(x - v_c t)dx$$

$$= \mu[(u_1 - u_0)v_p + u_0 v_c] + \mu[1 - (v_p/v_c)^2]v_0 v_c t. \quad (53)$$

The rate of flow of mechanical momentum into the rod, namely, $\partial_t p_x(t)$, is thus seen to be precisely equal to the total optical force $F_x$ exerted by the leading edge of the light pulse; see Eq.(52). The small constant term on the right-hand side of Eq.(53) accounts for the momentum picked up by the rod during the early stages of entry, when the leading-edge of the light pulse arrives at the entrance facet.

Note that the displacement $u(x,t)$ depicted in Fig.15 is the superposition of two functions, one that travels with the speed of light, $v_c$, and another that travels behind the first one at the speed of sound, $v_p$. Building on this idea, we may plot $u(x,t)$ at an instant of time when the pulse, whose duration is $T$, has fully entered the rod. In Fig.16, which corresponds to $t_0 > T$, the positive displacement represents that part of the function $u(x,t)$ which moves forward at the speed of light, $v_c$, whereas the negative displacement represents the part that travels along $x$ at the speed of sound, $v_p$.

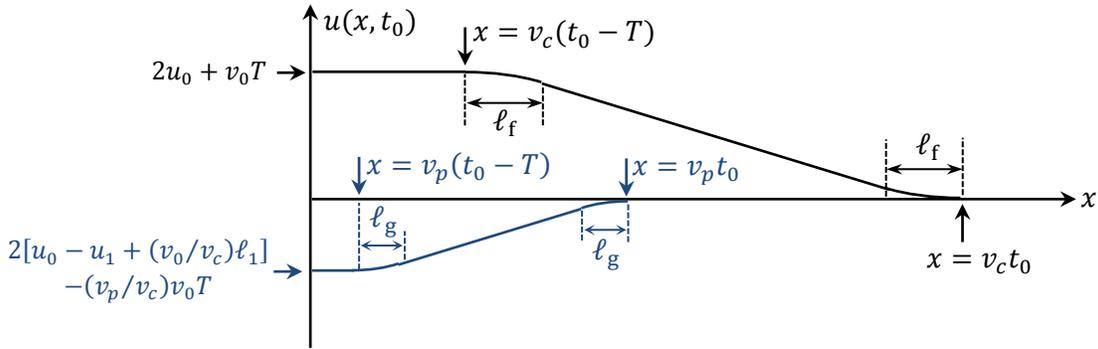

**Fig.16**. Two contributions to the displacement $u(x,t)$ at an instant of time $t_0$ when the light pulse of duration $T$ is fully inside the rod. The positive (upper) displacement travels along the $x$-axis at the speed of light, $v_c$, with the negative (lower) displacement trailing behind at the speed of sound, $v_p$.

Suppose now that the entrance facet of the rod in Fig.15 is antireflection coated, so that a net force of $-F_0 \hat{x}$ pulls on the front facet, while the leading edge of the pulse travels inside the rod with velocity $v_c$, as before. Figure 17 shows that the general form of $u(x,t)$ will now differ from that given by Eq.(51) only when $0 \leq x \leq v_p t$. Specifically, the linear segment of $u(x,t)$ in the interval $0 \leq x \leq v_p t - \ell_g$ is now given by $u_1 + (1 - v_p/v_c)v_0 t + (F_0/\sigma)(x - v_p t)$, and the left-hand boundary conditions of $g(x)$ are $g(-\ell_g) = u_1 - u_0 - (F_0/\sigma)\ell_g$ and $g'(-\ell_g) = F_0/\sigma$.

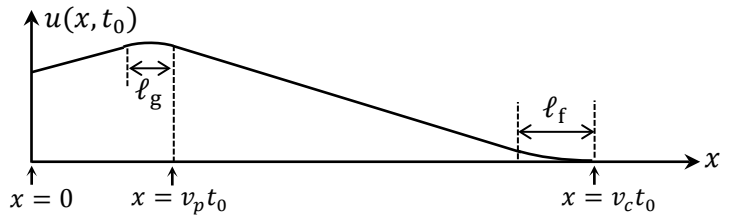

**Fig.17**. When the entrance facet of the rod in Fig.15 is antireflection coated, a constant force $-F_0 \hat{x}$ pulls on that facet, causing the displacement function $u(x,t)$ to acquire a positive slope, $F_0/\sigma$, in the interval between $x = 0$ and $x = v_p t - \ell_g$.

The mechanical momentum of the slab at time $t$ is now given by

$$p_x(t) = \int_0^\infty \mu \partial_t u(x,t) dx = \mu[(1 - v_p/v_c)v_0 - v_p(F_0/\sigma)](v_p t - \ell_g)$$
$$+ \int_{v_p t - \ell_g}^{v_p t} \mu[(1 - v_p/v_c)v_0 - v_p g'(x - v_p t)]dx + \mu v_0[(v_c - v_p)t - \ell_f]$$
$$- \mu v_c \int_{v_c t - \ell_f}^{v_c t} f'(x - v_c t)dx$$
$$= \mu[(u_1 - u_0)v_p + u_0 v_c] + (F_x - F_0)t. \tag{54}$$

As expected, the influx of mechanical momentum into the rod is somewhat reduced by the pull of the light at the antireflection coating layer. Nevertheless, the overall strength of the EM field traveling inside the rod is now greater as a result of the presence of the antireflection coating at the front facet, causing the magnitude of the force $F_x$ at the leading edge of the pulse to be greater than before.

**7. Elastic vibrations of a one-dimensional inhomogeneous slab**. In this section we generalize the results of Sec. 2 to cover the case of inhomogeneous slabs in one dimension. Consider the one-dimensional periodic medium depicted in Fig.18, where the mass-density and stiffness coefficient in the interval $0 \le x < d_1$ are $(\mu_1, \sigma_1)$, while the corresponding parameters in the interval $d_1 \le x < d_1 + d_2$ are $(\mu_2, \sigma_2)$. Defining $k_1 = \sqrt{\mu_1/\sigma_1}\,\omega$ and $k_2 = \sqrt{\mu_2/\sigma_2}\,\omega$ (both positive), a separable solution for the displacement $u(x,t)$ may be written as follows:

$$\begin{cases} u_1(x,t) = [A_1 \exp(ik_1 x) + B_1 \exp(-ik_1 x)] \exp(-i\omega t); & (0 \le x \le d_1), \\ u_2(x,t) = [A_2 \exp(ik_2 x) + B_2 \exp(-ik_2 x)] \exp(-i\omega t); & (d_1 \le x \le d_1 + d_2). \end{cases} \tag{55}$$

Here $\omega$, an arbitrarily chosen, real-valued, positive constant represents the temporal vibration frequency of the phonon under consideration. At $x = d_1$, the boundary conditions are

$$u_1(d_1, t) = u_2(d_1, t), \tag{56a}$$
$$\sigma_1 \partial_x u_1(d_1, t) = \sigma_2 \partial_x u_2(d_1, t). \tag{56b}$$

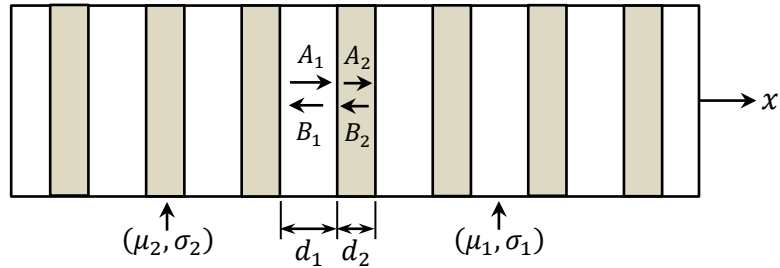

**Fig.18**. An infinitely long, periodic slab of material consists of repeated segments having widths $d_1$ and $d_2$. The mass-density and stiffness coefficient of these segments are $(\mu_1, \sigma_1)$ and $(\mu_2, \sigma_2)$, respectively.

Invoking Eq.(55), the above boundary conditions may be written

$$A_1 \exp(ik_1 d_1) + B_1 \exp(-ik_1 d_1) = A_2 \exp(ik_2 d_1) + B_2 \exp(-ik_2 d_1), \tag{57a}$$
$$k_1 \sigma_1 [A_1 \exp(ik_1 d_1) - B_1 \exp(-ik_1 d_1)] = k_2 \sigma_2 [A_2 \exp(ik_2 d_1) - B_2 \exp(-ik_2 d_1)]. \tag{57b}$$

Solving Eqs.(57) for $(A_2, B_2)$ in terms of $(A_1, B_1)$, we find

$$\begin{pmatrix} \exp(ik_2 d_1) & \exp(-ik_2 d_1) \\ k_2\sigma_2 \exp(ik_2 d_1) & -k_2\sigma_2 \exp(-ik_2 d_1) \end{pmatrix} \begin{pmatrix} A_2 \\ B_2 \end{pmatrix} = \begin{pmatrix} A_1 \exp(ik_1 d_1) + B_1 \exp(-ik_1 d_1) \\ k_1\sigma_1 [A_1 \exp(ik_1 d_1) - B_1 \exp(-ik_1 d_1)] \end{pmatrix}$$

$$\rightarrow \begin{pmatrix} A_2 \\ B_2 \end{pmatrix} = \tfrac{1}{2} \begin{pmatrix} \exp(-ik_2 d_1) & (k_2\sigma_2)^{-1} \exp(-ik_2 d_1) \\ \exp(ik_2 d_1) & -(k_2\sigma_2)^{-1} \exp(ik_2 d_1) \end{pmatrix} \begin{pmatrix} A_1 \exp(ik_1 d_1) + B_1 \exp(-ik_1 d_1) \\ k_1\sigma_1 [A_1 \exp(ik_1 d_1) - B_1 \exp(-ik_1 d_1)] \end{pmatrix}$$

$$= \tfrac{1}{2} \begin{pmatrix} A_1[1 + (k_1\sigma_1/k_2\sigma_2)] \exp[i(k_1 - k_2)d_1] + B_1[1 - (k_1\sigma_1/k_2\sigma_2)] \exp[-i(k_1 + k_2)d_1] \\ A_1[1 - (k_1\sigma_1/k_2\sigma_2)] \exp[i(k_1 + k_2)d_1] + B_1[1 + (k_1\sigma_1/k_2\sigma_2)] \exp[-i(k_1 - k_2)d_1] \end{pmatrix}. \quad (58)$$

At $x = d_1 + d_2$, the boundary conditions are

$$u_2(d_1 + d_2, t) = \eta u_1(0, t), \quad (59a)$$

$$\sigma_2 \partial_x u_2(d_1 + d_2, t) = \eta \sigma_1 \partial_x u_1(0, t). \quad (59b)$$

Here $\eta$ is an arbitrary constant, which must be subsequently determined. Invoking Eq.(55), the boundary conditions in Eq.(59) are written

$$A_2 \exp[ik_2(d_1 + d_2)] + B_2 \exp[-ik_2(d_1 + d_2)] = \eta(A_1 + B_1), \quad (60a)$$

$$k_2\sigma_2 \{A_2 \exp[ik_2(d_1 + d_2)] - B_2 \exp[-ik_2(d_1 + d_2)]\} = k_1\sigma_1 \eta (A_1 - B_1). \quad (60b)$$

Solving Eqs.(60) for $(A_1, B_1)$ in terms of $(A_2, B_2)$, we obtain

$$\eta \begin{pmatrix} 1 & 1 \\ k_1\sigma_1 & -k_1\sigma_1 \end{pmatrix} \begin{pmatrix} A_1 \\ B_1 \end{pmatrix} = \begin{pmatrix} \exp[ik_2(d_1 + d_2)] & \exp[-ik_2(d_1 + d_2)] \\ k_2\sigma_2 \exp[ik_2(d_1 + d_2)] & -k_2\sigma_2 \exp[-ik_2(d_1 + d_2)] \end{pmatrix} \begin{pmatrix} A_2 \\ B_2 \end{pmatrix}$$

$$\rightarrow \eta \begin{pmatrix} A_1 \\ B_1 \end{pmatrix} = \tfrac{1}{2} \begin{pmatrix} 1 & 1/(k_1\sigma_1) \\ 1 & -1/(k_1\sigma_1) \end{pmatrix} \begin{pmatrix} \exp[ik_2(d_1 + d_2)] & \exp[-ik_2(d_1 + d_2)] \\ k_2\sigma_2 \exp[ik_2(d_1 + d_2)] & -k_2\sigma_2 \exp[-ik_2(d_1 + d_2)] \end{pmatrix} \begin{pmatrix} A_2 \\ B_2 \end{pmatrix}$$

$$\rightarrow \eta \begin{pmatrix} A_1 \\ B_1 \end{pmatrix} = \tfrac{1}{2} \begin{pmatrix} [1 + (k_2\sigma_2/k_1\sigma_1)] \exp[ik_2(d_1 + d_2)] & [1 - (k_2\sigma_2/k_1\sigma_1)] \exp[-ik_2(d_1 + d_2)] \\ [1 - (k_2\sigma_2/k_1\sigma_1)] \exp[ik_2(d_1 + d_2)] & [1 + (k_2\sigma_2/k_1\sigma_1)] \exp[-ik_2(d_1 + d_2)] \end{pmatrix} \begin{pmatrix} A_2 \\ B_2 \end{pmatrix}. \quad (61)$$

Next, we define the parameter $\rho = k_1\sigma_1/k_2\sigma_2 = \sqrt{\mu_1\sigma_1/\mu_2\sigma_2}$, and substitute from Eq.(58) into Eq.(61) to arrive at

$$\eta \begin{pmatrix} A_1 \\ B_1 \end{pmatrix} = \tfrac{1}{4} \begin{pmatrix} (1 + \rho^{-1}) \exp[ik_2(d_1 + d_2)] & (1 - \rho^{-1}) \exp[-ik_2(d_1 + d_2)] \\ (1 - \rho^{-1}) \exp[ik_2(d_1 + d_2)] & (1 + \rho^{-1}) \exp[-ik_2(d_1 + d_2)] \end{pmatrix}$$

$$\times \begin{pmatrix} A_1(1 + \rho) \exp[i(k_1 - k_2)d_1] + B_1(1 - \rho) \exp[-i(k_1 + k_2)d_1] \\ A_1(1 - \rho) \exp[i(k_1 + k_2)d_1] + B_1(1 + \rho) \exp[-i(k_1 - k_2)d_1] \end{pmatrix}$$

$$= \tfrac{1}{4} \begin{pmatrix} A_1(2 + \rho + \rho^{-1}) \exp[i(k_1 d_1 + k_2 d_2)] - B_1(\rho - \rho^{-1}) \exp[-i(k_1 d_1 - k_2 d_2)] \\ +A_1(2 - \rho - \rho^{-1}) \exp[i(k_1 d_1 - k_2 d_2)] + B_1(\rho - \rho^{-1}) \exp[-i(k_1 d_1 + k_2 d_2)] \\ \\ A_1(\rho - \rho^{-1}) \exp[i(k_1 d_1 + k_2 d_2)] + B_1(2 - \rho - \rho^{-1}) \exp[-i(k_1 d_1 - k_2 d_2)] \\ -A_1(\rho - \rho^{-1}) \exp[i(k_1 d_1 - k_2 d_2)] + B_1(2 + \rho + \rho^{-1}) \exp[-i(k_1 d_1 + k_2 d_2)] \end{pmatrix}. \quad (62)$$

From Eq.(62), the ratio $A_1/B_1$ is seen to satisfy the following equation:

$$\frac{A_1}{B_1} = \frac{(A_1/B_1)[2\cos(k_2 d_2)+i(\rho+\rho^{-1})\sin(k_2 d_2)]\exp(ik_1 d_1) - i(\rho-\rho^{-1})\sin(k_2 d_2)\exp(-ik_1 d_1)}{(A_1/B_1)[i(\rho-\rho^{-1})\sin(k_2 d_2)]\exp(ik_1 d_1) + [2\cos(k_2 d_2)-i(\rho+\rho^{-1})\sin(k_2 d_2)]\exp(-ik_1 d_1)}. \quad (63)$$

This yields a quadratic equation for $A_1/B_1$, namely,

$$(\rho-\rho^{-1})\sin(k_2 d_2)\exp(ik_1 d_1)(A_1/B_1)^2 - 2[2\sin(k_1 d_1)\cos(k_2 d_2) + (\rho+\rho^{-1})\cos(k_1 d_1)\sin(k_2 d_2)](A_1/B_1)$$
$$+ (\rho-\rho^{-1})\sin(k_2 d_2)\exp(-ik_1 d_1) = 0. \quad (64)$$

Solving Eq.(64), we obtain two solutions for $A_1/B_1$, as follows:

$$\frac{A_1}{B_1} = \frac{2\sin(k_1 d_1)\cos(k_2 d_2)+(\rho+\rho^{-1})\cos(k_1 d_1)\sin(k_2 d_2) \pm \sqrt{[2\sin(k_1 d_1)\cos(k_2 d_2)+(\rho+\rho^{-1})\cos(k_1 d_1)\sin(k_2 d_2)]^2 - [(\rho-\rho^{-1})\sin(k_2 d_2)]^2}}{(\rho-\rho^{-1})\sin(k_2 d_2)\exp(ik_1 d_1)}. \quad (65)$$

Aside from the phase-factor $\exp(ik_1 d_1)$ in the denominator of Eq.(65), the two solutions thus obtained for $A_1/B_1$ are readily seen to be inverses of each other. If the expression under the radical happens to be positive, these two values for $A_1/B_1$ will result in a pair of conjugate solutions for $u(x,t)$ throughout the entire medium. Returning now to Eq.(62), we solve for $\eta$ and, with the aid of Eq.(65), we find

$$\eta = \tfrac{1}{2}i(A_1/B_1)(\rho-\rho^{-1})\sin(k_2 d_2)\exp(ik_1 d_1) + \cos(k_2 d_2)\exp(-ik_1 d_1) - \tfrac{1}{2}i(\rho+\rho^{-1})\sin(k_2 d_2)\exp(-ik_1 d_1)$$

$$\rightarrow \quad \eta = \cos(k_1 d_1)\cos(k_2 d_2) - \tfrac{1}{2}(\rho+\rho^{-1})\sin(k_1 d_1)\sin(k_2 d_2)$$
$$\pm\tfrac{1}{2}i\sqrt{[2\sin(k_1 d_1)\cos(k_2 d_2)+(\rho+\rho^{-1})\cos(k_1 d_1)\sin(k_2 d_2)]^2 - [(\rho-\rho^{-1})\sin(k_2 d_2)]^2}. \quad (66)$$

When the expression under the radical is positive, the complex number $\eta$ will have unit magnitude and may be written as $\eta = \exp(i\varphi)$; the phase $\varphi$, of course, depends on the chosen temporal frequency $\omega$, as well as on the material parameters $(\mu_1, \sigma_1, d_1)$ and $(\mu_2, \sigma_2, d_2)$. The real part of $\eta$ given by Eq.(66) will be equal to $\cos\varphi$, that is,

$$\cos\varphi = \frac{(\rho+1)^2 \cos(k_1 d_1 + k_2 d_2) - (\rho-1)^2 \cos(k_1 d_1 - k_2 d_2)}{4\rho}. \quad (67)$$

At $\omega = 0$, where $k_1 d_1 \pm k_2 d_2 = (\sqrt{\mu_1/\sigma_1}\, d_1 \pm \sqrt{\mu_2/\sigma_2}\, d_2)\omega = 0$, we have $\cos\varphi = 1$. As $\omega$ increases, $\cos\varphi$ drops at first, but eventually it exits the interval $[-1, +1]$, at which point the expression under the radical in Eq.(66) becomes negative. The range of frequencies $\omega$ in which both values of $\eta$ are real represents a bandgap, within which long-range phonons cannot exist. As $\omega$ continues to climb, $\cos\varphi$ in Eq.(67) returns to the interval $[-1, +1]$, thus creating another frequency band for phonons. Each allowed band, of course, is followed by a band-gap and then by a new band of allowed phonon frequencies. Needless to say, the existence of the phase-angle $\varphi$ is what enables the definition of a Bloch wave-vector $\boldsymbol{k}$ for the system of Fig.18, where $\varphi = k(d_1 + d_2)$.

In the special case when $\rho = 1$, that is, $\mu_1\sigma_1 = \mu_2\sigma_2$, the only solution of Eq.(64) turns out to be $A_1/B_1 = 0$. Even though $k_1$ could still differ from $k_2$, the overall solution of the problem is trivial, and, as Eq.(67) reveals, there will be no bandgaps.

When the expression under the radical in Eqs.(65) and (66) is negative, $A_1/B_1$ becomes a complex number of unit magnitude, having a phase-angle that depends on $\omega$ as well as on material parameters. At the same time, the two values of $\eta$ become real, one greater and the other less than unity. These values of $A_1/B_1$ and $\eta$ represent two distinct solutions for $u_\omega(x)$, one growing and the other declining exponentially along the $x$-axis. Each of these eigen-functions has its own complex conjugate, representing two additional solutions associated with negative $\omega$. Note, with reference to Eqs.(55), (58) and (65), that a simultaneous reversal of the signs of $\omega$, $k_1$ and $k_2$ yields the complex conjugate solution $u_\omega^*(x)$ *without* changing the corresponding value of $\eta$ given by Eq.(66). In practice, one ignores the values of $\omega$ that lead to exponentially growing or decaying eigen-functions $u_\omega(x)$, relegating them to "band-gaps" in the frequency domain.

**7.1. Finite-width periodic slab**. The Bloch solutions of the wave equation associated with the phase-factor $\eta = \exp(i\varphi) = \exp[ik(d_1 + d_2)]$ may be written as $u_\omega(x,t) = u_\omega(x)\exp[i(kx - \omega t)]$, where the (generally complex) function $u_\omega(x)$ is periodic, having a periodicity of $d_1 + d_2$. These are the same solutions as in Eq.(55), except that the linear phase-factor $\exp(ikx)$ has been factored out in order to emphasize the fact that $u_\omega(x)$ is periodic. Considering that the sign of $\omega$ is immaterial — in the sense that $u_\omega(x)\exp(ikx)$ and its conjugate $u_\omega^*(x)\exp(-ikx)$ are equally valid spatial profiles satisfying the wave equation irrespective of whether $\omega$ is positive or negative — we conclude that separable solutions of the wave equation may be equivalently written in the following form:

$$u_\omega(x,t) = |u_\omega(x)|\{\Lambda_1 \sin[kx + \psi_\omega(x)] + \Lambda_2 \cos[kx + \psi_\omega(x)]\}\{\Lambda_3 \sin\omega t + \Lambda_4 \cos\omega t\}$$

$$= \Lambda |u_\omega(x)| \sin[kx + \psi_\omega(x) + \phi_0] \sin(\omega t + \phi_1). \qquad (68)$$

In the above equation, $u_\omega(x) = |u_\omega(x)|\exp[i\psi_\omega(x)]$ is periodic, $k$ is a function of $\omega$, and the real-valued coefficients $\Lambda_1, \Lambda_2, \Lambda_3, \Lambda_4, \Lambda, \phi_0, \phi_1$ are arbitrary constants. If a boundary, say, that at $x = -L$, is immobile, then $\phi_0$ must be chosen to ensure that $u_\omega(-L, t) = 0$. If the other boundary at $x = L$ is also immobile, then only those values of $\omega$ will be acceptable that satisfy $\sin[kL + \psi_\omega(L) + \phi_0] = 0$. Similarly, if the boundaries at $x = \pm L$ happen to be free, then $\phi_0$ and $\omega$ must be chosen to ensure that $\partial_x u_\omega(\pm L, t) = 0$. In this way, a subset of values of $\omega$ is selected for which $u_\omega(x,t)$ of Eq.(68) is a Bloch wave that satisfies the wave equation as well as the specified boundary conditions. Needless to say, all such values of $\omega$ are positive, since the negative values of $\omega$ are automatically incorporated into the eigen-functions defined by Eq.(68).

The initial conditions $u(x, t = 0)$ and $v(x, t = 0)$ may now be expanded in terms of the spatial parts of the eigen-functions given by Eq.(68) which also satisfy the boundary conditions, namely,

$$u(x, t = 0) = \sum_\omega \Lambda_\omega |u_\omega(x)| \sin[kx + \psi_\omega(x) + \phi_{0\omega}], \qquad (69a)$$

$$v(x, t = 0) = \sum_\omega \tilde{\Lambda}_\omega |u_\omega(x)| \sin[kx + \psi_\omega(x) + \phi_{0\omega}]. \qquad (69b)$$

Consequently, the general solution of the wave equation is given by

$$u(x,t) = \sum_\omega |u_\omega(x)| \sin[kx + \psi_\omega(x) + \phi_{0\omega}] [\Lambda_\omega \cos(\omega t) + \tilde{\Lambda}_\omega \sin(\omega t)/\omega]. \qquad (70)$$

Note in the above equations that $u_\omega(x)$ is dimensionless, whereas $\Lambda_\omega$ and $\tilde{\Lambda}_\omega$ have the units of length and velocity, respectively. The overall linear momentum $p(t)$ of a slab of width $2L$ is readily derived from its mass and velocity profiles, as follows:

$$p(t) = \int_{-L}^{L} \mu(x) \partial_t u(x,t) dx$$

$$= \sum_\omega [\tilde{\Lambda}_\omega \cos(\omega t) - \Lambda_\omega \omega \sin(\omega t)] \int_{-L}^{L} \mu(x)|u_\omega(x)| \sin[kx + \psi_\omega(x) + \phi_{0\omega}] \, dx. \quad (71)$$

If the slab happens to be free at both ends, the integral appearing in Eq.(71) will identically vanish for all nonzero frequencies $\omega$ — this is readily proven by integrating the wave equation $\partial_x[\sigma(x)\partial_x u_\omega(x,t)] = -\omega^2 \mu(x) u_\omega(x,t)$, satisfied by all eigen-functions in Eq.(68), over the interval $[-L, L]$. The only surviving term in Eq.(71) corresponds to $\omega = 0$, whose eigen-function is $u_0(x,t) = 1$. Consequently, $p(t) = \tilde{\Lambda}_0 \int_{-L}^{L} \mu(x) dx$, where $\tilde{\Lambda}_0$ is the coefficient of the zero-frequency term in Eq.(69b), while $\int_{-L}^{L} \mu(x) dx$ is the total mass of the slab. Note that $\tilde{\Lambda}_0$ is *not* necessarily the average initial velocity of the slab, as the eigen-functions $u_\omega(x,t)$ may not have a vanishing integral over the interval $[-L, L]$ when $\omega \neq 0$.

**7.2. Eigen-function orthogonality**. In general, an inhomogeneous one-dimensional medium is specified by its mass-density $\mu(x)$ and stiffness coefficient $\sigma(x)$, both of which vary with position along the $x$-axis. In such a medium, the wave equation governing the displacement $u(x,t)$ is derived from Newton's second law of motion, as follows:

$$\partial_x[\sigma(x)\partial_x u(x,t)] = \mu(x)\partial_t^2 u(x,t). \quad (72)$$

The eigen-modes of the above equation are obtained by separation of variables in the form of $u(x,t) = u(x) \exp(\pm i\omega t)$, which leads to

$$\sigma(x)\ddot{u}(x) + \dot{\sigma}(x)\dot{u}(x) + \omega^2 \mu(x) u(x) = 0. \quad (73)$$

Here differentiation with respect to $x$ is indicated by an over-dot. Note that $u(x)$, a complex function in general, does not depend on the choice of sign for $\omega$. However, since $u^*(x)$, the conjugate of $u(x)$, is also a solution of Eq.(73) — because $\mu(x)$, $\sigma(x)$, and $\omega$ are real — we may arbitrarily assign $u(x)$ to positive $\omega$, in which case $u^*(x)$ gets automatically assigned to $-\omega$.

Suppose now that $u_{\omega 1}(x)$ and $u_{\omega 2}(x)$ are two eigen-functions of Eq.(73), corresponding respectively to distinct temporal frequencies $\omega_1$ and $\omega_2$. We will have

$$\sigma(x)\ddot{u}_{\omega 1}(x) + \dot{\sigma}(x)\dot{u}_{\omega 1}(x) + \omega_1^2 \mu(x) u_{\omega 1}(x) = 0, \quad (74a)$$

$$\sigma(x)\ddot{u}_{\omega 2}(x) + \dot{\sigma}(x)\dot{u}_{\omega 2}(x) + \omega_2^2 \mu(x) u_{\omega 2}(x) = 0. \quad (74b)$$

Multiplying the first of the above equations by $u_{\omega 2}(x)$ and the second by $u_{\omega 1}(x)$, then subtracting one equation from the other, yields

$$u_{\omega 2}(x)\frac{d}{dx}[\sigma(x)\dot{u}_{\omega 1}(x)] - u_{\omega 1}(x)\frac{d}{dx}[\sigma(x)\dot{u}_{\omega 2}(x)] + (\omega_1^2 - \omega_2^2)\mu(x) u_{\omega 1}(x) u_{\omega 2}(x) = 0. \quad (75)$$

Next, we integrate Eq.(75) over the interval $[a, b]$ and apply the method of integration by parts to arrive at

$$[\sigma(b)\dot{u}_{\omega 1}(b)u_{\omega 2}(b) - \sigma(a)\dot{u}_{\omega 1}(a)u_{\omega 2}(a)] - \int_a^b \sigma(x)\dot{u}_{\omega 1}(x)\dot{u}_{\omega 2}(x)dx$$

$$-[\sigma(b)u_{\omega 1}(b)\dot{u}_{\omega 2}(b) - \sigma(a)u_{\omega 1}(a)\dot{u}_{\omega 2}(a)] + \int_a^b \sigma(x)\dot{u}_{\omega 1}(x)\dot{u}_{\omega 2}(x)dx$$

$$+(\omega_1^2 - \omega_2^2)\int_a^b \mu(x)u_{\omega 1}(x)u_{\omega 2}(x)dx = 0. \quad (76)$$

If the boundary conditions are such that the bracketed terms in Eq.(76) vanish (or cancel out), and if $|\omega_1| \neq |\omega_2|$, we observe that $u_{\omega 1}(x)$ and $u_{\omega 2}(x)$, over the interval $[a, b]$ and in the presence of the weight function $\mu(x)$, are orthogonal to each other. In other words,

$$\int_a^b \mu(x) u_{\omega 1}(x) u_{\omega 2}(x) dx = 0. \tag{77}$$

A few words must be said at this point about the zero-frequency eigen-function, which is readily obtained from Eq.(72) as $u_0(x, t) = c_1 + c_2 \int_a^x \sigma^{-1}(x') dx'$; here $c_1$ and $c_2$ are arbitrary (real-valued) constants. Considering that $\sigma(x) > 0$ throughout the interval $[a, b]$, the usual boundary conditions, $u_0(a, t) = 0$, $u_0(b, t) = 0$, $\partial_x u_0(a, t) = 0$, or $\partial_x u_0(b, t) = 0$, ensure that $c_2 = 0$. Thus, when one or both boundaries are immobile, $u_0(x, t) = 0$, whereas $u_0(x, t) = c_1$ when both boundaries are free.

The initial position $u(x, t = 0)$ and velocity $v(x, t = 0)$ of the medium may now be expanded in terms of the above eigen-functions $u_\omega(x)$, as follows:

$$u(x, t = 0) = \int_{-\infty}^{\infty} U_0(\omega) u_\omega(x) d\omega, \tag{78a}$$

$$v(x, t = 0) = \int_{-\infty}^{\infty} V_0(\omega) u_\omega(x) d\omega. \tag{78b}$$

Note that some values of $\omega$ (e.g., those within a bandgap) may not possess acceptable eigen-functions, in which case the domain of integration in Eq.(78) will be a subset of the real axis. Considering that the integrals in Eq.(78) range over both positive and negative values of $\omega$, and that $u_{-\omega}(x) = u_\omega^*(x)$, the functions $U_0(\omega)$ and $V_0(\omega)$ must be Hermitian. Since $u_\omega(x)$ is not necessarily orthogonal to $u_\omega^*(x)$, care must be taken when determining the functions $U_0(\omega)$ and $V_0(\omega)$; this problem is addressed in some detail in Appendix C. Given that each eigen-function $u_\omega(x)$ is associated with both temporal functions $\exp(i\omega t)$ and $\exp(-i\omega t)$, we must write

$$u(x, t) = \int_{-\infty}^{\infty} [U_+(\omega) u_\omega(x) \exp(i\omega t) + U_-(\omega) u_\omega(x) \exp(-i\omega t)] d\omega, \tag{79a}$$

$$v(x, t) = \partial_t u(x, t) = \int_{-\infty}^{\infty} i\omega [U_+(\omega) u_\omega(x) \exp(i\omega t) - U_-(\omega) u_\omega(x) \exp(-i\omega t)] d\omega. \tag{79b}$$

A quick glance at Eqs.(78) and (79) now yields the functions $U_\pm(\omega)$, as follows:

$$U_\pm(\omega) = \tfrac{1}{2}[U_0(\omega) \mp i V_0(\omega)/\omega]. \tag{80}$$

Finally, substitution from Eq.(80) into Eqs.(79) yields

$$u(x, t) = \int_{-\infty}^{\infty} [U_0(\omega) \cos(\omega t) + \omega^{-1} V_0(\omega) \sin(\omega t)] u_\omega(x) d\omega, \tag{81a}$$

$$v(x, t) = \int_{-\infty}^{\infty} [V_0(\omega) \cos(\omega t) - \omega U_0(\omega) \sin(\omega t)] u_\omega(x) d\omega. \tag{81b}$$

These are general formulas for computing the position and velocity profiles of the medium as functions of time starting with the initial conditions $u(x, t = 0)$ and $v(x, t = 0)$.

**7.3. Momentum conservation**. Integrating the fundamental Eq.(73) over the $[a, b]$ interval, we find

$$\omega^2 \int_a^b \mu(x) u_\omega(x) dx = \sigma(a) \dot{u}_\omega(a) - \sigma(b) \dot{u}_\omega(b). \tag{82}$$

For the overall momentum of the system to be conserved, the boundaries at $x = a$ and $x = b$ must be free, that is, $\dot{u}_\omega(a) = \dot{u}_\omega(b) = 0$. Therefore,

$$\int_a^b \mu(x) u_\omega(x) dx = 0, \qquad (\omega \neq 0). \tag{83}$$

In the special case of $\omega = 0$, the integral $\int_a^b \mu(x) u_0(x) dx$ can be nonzero. In fact, $u_0(x)$ should be a real-valued constant, as this choice satisfies both Eq.(73) and the free boundary conditions. Next, we invoke Eq.(78b) to write

$$\int_a^b \mu(x) v(x, t = 0) u_0(x) dx$$
$$= \int_{-\infty}^\infty V_0(\omega) \int_a^b \mu(x) u_\omega(x) u_0(x) dx\, d\omega = V_0(0) \int_a^b \mu(x) u_0^2(x) dx. \tag{84}$$

Finally, with the aid of Eqs.(81b), (83), and (84), the overall momentum of the system at time $t$ may be evaluated as follows:

$$p(t) = \int_a^b \mu(x) v(x,t) dx = \int_{-\infty}^\infty [V_0(\omega)\cos(\omega t) - \omega U_0(\omega)\sin(\omega t)] \int_a^b \mu(x) u_\omega(x) dx\, d\omega$$
$$= V_0(0) \int_a^b \mu(x) u_0(x) dx = \left[\frac{\int_a^b \mu(x) u_0(x) dx}{\int_a^b \mu(x) u_0^2(x) dx}\right] \int_a^b \mu(x) v(x, t = 0) u_0(x) dx. \tag{85}$$

Considering that the time $t$ has disappeared from the above equation, and that $u_0(x)$ is a constant over the $[a, b]$ interval, Eq.(85) confirms that the overall momentum of the system is conserved.

**7.4. Energy conservation**. Invoking Eq.(81b) and the orthogonality condition, Eq.(77), the kinetic energy of the system at time $t$ may be expressed as follows:

$$\mathcal{E}_K(t) = \tfrac{1}{2} \int_a^b \mu(x) v^2(x, t) dx$$
$$= \tfrac{1}{2} \iint_{-\infty}^\infty [V_0(\omega)\cos(\omega t) - \omega U_0(\omega)\sin(\omega t)][V_0(\omega')\cos(\omega' t) - \omega' U_0(\omega')\sin(\omega' t)] \int_a^b \mu(x) u_\omega(x) u_{\omega'}(x) dx\, d\omega\, d\omega'$$
$$= \tfrac{1}{2} \int_{-\infty}^\infty [V_0(\omega)\cos(\omega t) - \omega U_0(\omega)\sin(\omega t)]^2 \int_a^b \mu(x) u_\omega^2(x) dx\, d\omega$$
$$+ \tfrac{1}{2} \int_{-\infty}^\infty [V_0(\omega)\cos(\omega t) - \omega U_0(\omega)\sin(\omega t)][V_0^*(\omega)\cos(\omega t) - \omega U_0^*(\omega)\sin(\omega t)] \int_a^b \mu(x) u_\omega(x) u_\omega^*(x) dx\, d\omega. \tag{86}$$

As for the potential energy, we first need to prove the following identity using the fundamental Eq.(73) and the method of integration by parts:

$$\int_a^b \sigma(x)\dot{u}_\omega(x)\dot{u}_{\omega'}(x) dx = [\sigma(b)\dot{u}_\omega(b)u_{\omega'}(b) - \sigma(a)\dot{u}_\omega(a)u_{\omega'}(a)]^{\,0} - \int_a^b \tfrac{d}{dx}[\sigma(x)\dot{u}_\omega(x)] u_{\omega'}(x) dx = \omega^2 \int_a^b \mu(x) u_\omega(x) u_{\omega'}(x) dx. \tag{87}$$

The bracketed term is the same one that appeared in Eq.(76), and has been set to zero because of the imposed boundary conditions. The potential energy of the system at time $t$ may now be evaluated using Eq.(81a), Eq.(87), and the orthogonality condition, Eq.(77), as follows:

$$\mathcal{E}_P(t) = \tfrac{1}{2} \int_a^b \sigma(x) \dot{u}^2(x, t) dx$$
$$= \tfrac{1}{2} \iint_{-\infty}^\infty [U_0(\omega)\cos(\omega t) + \omega^{-1} V_0(\omega)\sin(\omega t)][U_0(\omega')\cos(\omega' t) + \omega'^{-1} V_0(\omega')\sin(\omega' t)] \int_a^b \sigma(x)\dot{u}_\omega(x)\dot{u}_{\omega'}(x) dx\, d\omega\, d\omega'$$

$$= \tfrac{1}{2} \int_{-\infty}^{\infty} [U_0(\omega)\cos(\omega t) + \omega^{-1}V_0(\omega)\sin(\omega t)]^2 \omega^2 \int_a^b \mu(x) u_\omega^2(x) dx d\omega$$

$$+ \tfrac{1}{2} \int_{-\infty}^{\infty} [U_0(\omega)\cos(\omega t) + \omega^{-1}V_0(\omega)\sin(\omega t)][U_0^*(\omega)\cos(\omega t) + \omega^{-1}V_0^*(\omega)\sin(\omega t)] \omega^2 \int_a^b \mu(x) u_\omega(x) u_\omega^*(x) dx d\omega. \quad (88)$$

The total energy of the system is thus given by

$$\mathcal{E}_K(t) + \mathcal{E}_P(t) = \tfrac{1}{2} \int_{-\infty}^{\infty} [V_0^2(\omega) + \omega^2 U_0^2(\omega)] \int_a^b \mu(x) u_\omega^2(x) dx d\omega$$

$$+ \tfrac{1}{2} \int_{-\infty}^{\infty} [V_0(\omega)V_0^*(\omega) + \omega^2 U_0(\omega)U_0^*(\omega)] \int_a^b \mu(x) u_\omega(x) u_\omega^*(x) dx d\omega. \quad (89)$$

The fact that the sum of kinetic and potential energies in Eq.(89) is time-independent is proof that the total energy of the system is conserved. Note that, unlike momentum conservation, the conservation of energy proven here holds for both free and immobile boundary conditions.

**8. Concluding remarks**. We have shown that any momentum and kinetic energy deposited (e.g., by a short pulse of light) into a solid elastic medium, will evolve with time in such a way as to preserve the initial momentum and energy, despite the fact that the vibrational modes of the medium with non-zero frequency (i.e., $\omega \neq 0$) do not carry any momentum of their own. In other words, the entire mechanical momentum acquired by the medium is carried by the $\omega = 0$ vibrational mode. Also, in some instances in practice, the optical energy absorbed by the material medium is converted to kinetic energy of motion and/or thermal energy of elastic vibrations. In other instances, the exchanged optical energy may arise from the Doppler-shift of the reflected and/or transmitted photons. We have ignored the effects of such Doppler shifts in our analysis—see, for instance, the examples of Sec.4, where the light beam is either reflected from a mirror or transmitted through a dielectric slab. So long as the optical energy exchanged with matter is relatively small, the neglect of the Doppler shift could be justified—but of course, in extreme circumstances, there may occur situations where such effects need to be taken into consideration.

# Appendix A

The one-dimensional Fourier transform operator $\mathcal{F}$ acting on the function $f(x)$ is defined as follows:

$$\mathcal{F}\{f(x)\} = \int_{-\infty}^{\infty} f(x) \exp(-ikx)\, dx = F(k). \tag{A1}$$

The inverse Fourier transform operator $\mathcal{F}^{-1}$ then yields

$$\mathcal{F}^{-1}\{F(k)\} = \frac{1}{2\pi}\int_{-\infty}^{\infty} F(k) \exp(ikx)\, dk = f(x). \tag{A2}$$

Some important theorems are easily proven using the above definitions. The scaling theorem, for instance, asserts that

$$\mathcal{F}\{f(ax)\} = \frac{1}{|a|} F(k/a); \qquad (a = \text{arbitrary real number}). \tag{A3}$$

According to the differentiation theorem,

$$\mathcal{F}\{f'(x)\} = ik F(k). \tag{A4}$$

The central-value theorem and its inverse are written

$$\int_{-\infty}^{\infty} f(x)\, dx = F(0). \tag{A5}$$

$$\int_{-\infty}^{\infty} F(k)\, dk = 2\pi f(0). \tag{A6}$$

The convolution theorem and its inverse, the multiplication theorem, read as follows:

$$\mathcal{F}\{f(x) * g(x)\} = F(k)G(k), \tag{A7}$$

$$\mathcal{F}\{f(x)g(x)\} = \frac{1}{2\pi} F(k) * G(k). \tag{A8}$$

Here the convolution operation is defined as

$$f(x) * g(x) = \int_{-\infty}^{\infty} f(x')g(x-x')\, dx'. \tag{A9}$$

Finally, Parseval's theorem reads as follows:

$$\int_{-\infty}^{\infty} f(x) g^*(x)\, dx = \frac{1}{2\pi} \int_{-\infty}^{\infty} F(k) G^*(k)\, dk. \tag{A10}$$

The following Fourier transform pairs are readily evaluated using straightforward integration:

$$f(x) = \exp(-a|x|) \quad \rightarrow \quad F(k) = \frac{2a}{a^2 + k^2}; \qquad (a > 0). \tag{A11}$$

$$f(x) = \frac{1}{1 + x^2} \quad \rightarrow \quad F(k) = \pi \exp(-|k|). \tag{A12}$$

$$f(x) = \cos(ax) \quad \rightarrow \quad F(k) = \pi[\delta(k+a) + \delta(k-a)]. \tag{A13}$$

$$f(x) = \arctan(x/a) \quad \rightarrow \quad F(k) = \frac{\pi \exp(-a|k|)}{ik}; \qquad (a > 0). \tag{A14}$$

To prove the last identity note that

$$\tan f(x) = x/a \quad \rightarrow \quad f'(x)[1 + \tan^2 f(x)] = 1/a \quad \rightarrow \quad f'(x) = \frac{1/a}{1+(x/a)^2}. \tag{A15}$$

Now, with reference to Eq.(A12) and the scaling theorem, the Fourier transform of $f'(x)$ is $\pi \exp(-a|k|)$, which must be the same as $ikF(k)$ in accordance with the differentiation theorem. Consequently, $f(x) = \arctan(x/a)$ Fourier transforms to $F(k) = (\pi/ik)\exp(-a|k|)$.

Finally, we prove that the Fourier transform of $\text{comb}(x) = \sum_{n=-\infty}^{\infty} \delta(x-n)$ is the function $\text{comb}(k/2\pi)$. The following proof takes the scaled comb function $f(x) = (1/a)\,\text{comb}(x/a) = \sum_{n=-\infty}^{\infty} \delta(x - na)$, and proceeds to show that $F(k) = \lim_{N\to\infty} \sum_{n=-N}^{N} \exp(-ikna)$ is equal to $\text{comb}(ak/2\pi)$.

$$\sum_{n=-N}^{N} \exp(-inak) = \exp(iNak) \sum_{n=0}^{2N} \exp(-inak) = \exp(iNak)\left\{\frac{1-\exp[-i(2N+1)ak]}{1-\exp(-iak)}\right\}$$

$$= \frac{\exp[i(N+\tfrac12)ak] - \exp[-i(N+\tfrac12)ak]}{\exp(\tfrac12 iak) - \exp(-\tfrac12 iak)} = \frac{\sin[(N+\tfrac12)ak]}{\sin(\tfrac12 ak)}$$

> When $N\to\infty$, everything is negligible except the vicinity of $k = 2m\pi/a$, where the denominator vanishes.

$$\cong \sum_{m=-\infty}^{\infty} \frac{(-1)^m \sin[(N+\tfrac12)a(k-2m\pi/a)]}{\sin(m\pi) + \tfrac12 a\cos(m\pi)(k-2m\pi/a)}$$

$$= \sum_{m=-\infty}^{\infty} \frac{(-1)^m \sin[(N+\tfrac12)a(k-2m\pi/a)]}{\tfrac12 a(-1)^m (k-2m\pi/a)}$$

$$= \sum_{m=-\infty}^{\infty} \frac{2(N+\tfrac12)(a/\pi)}{(a/\pi)} \frac{\sin[(N+\tfrac12)a(k-2m\pi/a)]}{(N+\tfrac12)a(k-2m\pi/a)} \quad \boxed{\text{sinc}(k) = \sin(\pi k)/(\pi k).}$$

$$= \sum_{m=-\infty}^{\infty} (2\pi/a)(N+\tfrac12)(a/\pi)\,\text{sinc}[(N+\tfrac12)(a/\pi)(k-2m\pi/a)]$$

> sinc(·) approaches $\delta$-function when $N\to\infty$.

$$\cong \sum_{m=-\infty}^{\infty} (2\pi/a)\,\delta(k-2m\pi/a)$$

$$= \sum_{m=-\infty}^{\infty} \delta\left[\frac{a}{2\pi}\left(k-\frac{2m\pi}{a}\right)\right] = \sum_{m=-\infty}^{\infty} \delta\left(\frac{ak}{2\pi}-m\right) = \text{comb}\left(\frac{ak}{2\pi}\right). \tag{A16}$$

The proof is now complete.

# Appendix B

Shown in Fig. B1 is a semi-infinite substrate of refractive index $n_s$, coated with a dielectric layer of refractive index $n$ and thickness $d$. The normally-incident plane-wave of wavelength $\lambda_0$ is linearly-polarized along the $y$-axis, having electric and magnetic field amplitudes $E_0 \hat{y}$ and $H_0 \hat{z}$, where $H_0 = E_0/Z_0$. Here $Z_0 = \sqrt{\mu_0/\varepsilon_0}$ is the impedance of free space. Inside the dielectric layer, the forward-propagating beam has amplitude $E_1 \hat{y}$, whose magnitude relative to that of the incident beam is found to be

$$\frac{E_1}{E_0} = \frac{2/(1+n)}{1-[(n-n_s)/(n+n_s)][(n-1)/(n+1)]\exp(\mathrm{i}4\pi n d/\lambda_0)}. \tag{B1}$$

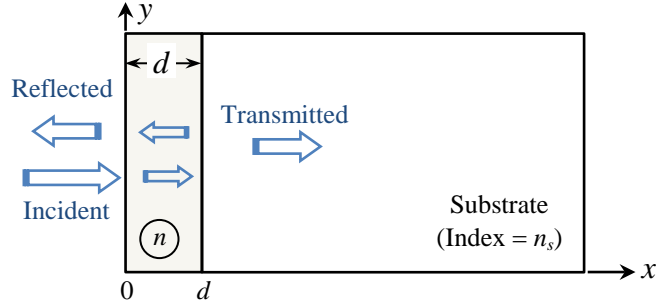

**Fig. B1**. Monochromatic, linearly-polarized plane-wave is normally incident on a dielectric layer of thickness $d$ and refractive index $n$, sitting atop a transparent, semi-infinite substrate of refractive index $n_s$. Also shown are the reflected wave, the counter-propagating waves within the dielectric layer, and the transmitted wave into the substrate. The incident beam's $E$-field amplitude is $E_0 \hat{y}$, while that of the plane-wave propagating to the right inside the dielectric layer is $E_1 \hat{y}$.

The total electric and magnetic fields inside the dielectric layer are readily found to be

$$E_y(x,t) = E_1 \left[ \exp\left(\frac{\mathrm{i}2\pi n x}{\lambda_0}\right) + \left(\frac{n-n_s}{n+n_s}\right) \exp\left(\frac{\mathrm{i}4\pi n d}{\lambda_0}\right) \exp\left(-\frac{\mathrm{i}2\pi n x}{\lambda_0}\right) \right] \exp(-\mathrm{i}\omega t), \tag{B2}$$

$$H_z(x,t) = \frac{nE_1}{Z_0} \left[ \exp\left(\frac{\mathrm{i}2\pi n x}{\lambda_0}\right) - \left(\frac{n-n_s}{n+n_s}\right) \exp\left(\frac{\mathrm{i}4\pi n d}{\lambda_0}\right) \exp\left(-\frac{\mathrm{i}2\pi n x}{\lambda_0}\right) \right] \exp(-\mathrm{i}\omega t). \tag{B3}$$

Considering that the polarization of the dielectric material is $\boldsymbol{P}(x,t) = \varepsilon_0 (n^2-1)\boldsymbol{E}(x,t)$, one may compute the time-averaged EM force-density acting on the dielectric layer as follows:

$$\langle \boldsymbol{f}(x,t) \rangle = \tfrac{1}{2}\mathrm{Re}[\partial_t \boldsymbol{P}(x,t) \times \mu_0 \boldsymbol{H}^*(x,t)] = \tfrac{1}{2}\mathrm{Im}[\omega \mu_0 \varepsilon_0 (n^2-1) E_y H_z^*] \hat{x}. \tag{B4}$$

In what follows, we shall treat several special cases of the above problem. As a first example, let us consider a free-standing layer of refractive index $n$ and thickness $d = \lambda_0/2n$ in the absence of the substrate. We will have

$$E_y(x,t) = E_0 \left(\frac{n+1}{2n}\right) \left[ \exp\left(\frac{\mathrm{i}2\pi n x}{\lambda_0}\right) + \left(\frac{n-1}{n+1}\right) \exp\left(-\frac{\mathrm{i}2\pi n x}{\lambda_0}\right) \right] \exp(-\mathrm{i}\omega t), \tag{B5}$$

$$H_z(x,t) = \frac{nE_0}{Z_0}\left(\frac{n+1}{2n}\right) \left[ \exp\left(\frac{\mathrm{i}2\pi n x}{\lambda_0}\right) - \left(\frac{n-1}{n+1}\right) \exp\left(-\frac{\mathrm{i}2\pi n x}{\lambda_0}\right) \right] \exp(-\mathrm{i}\omega t), \tag{B6}$$

$$\langle \boldsymbol{f}(x,t) \rangle = -\frac{\pi \varepsilon_0 E_0^2 (n^2-1)^2}{2n\lambda_0} \sin\left(\frac{4\pi n x}{\lambda_0}\right) \hat{x}. \tag{B7}$$

For our second example, we consider another free-standing dielectric layer of refractive index $n$ and thickness $d = \lambda_0/4n$, again without a substrate. We find

$$E_y(x,t) = E_0 \left(\frac{n+1}{n^2+1}\right) \left[\exp\left(\frac{i2\pi nx}{\lambda_0}\right) - \left(\frac{n-1}{n+1}\right) \exp\left(-\frac{i2\pi nx}{\lambda_0}\right)\right] \exp(-i\omega t), \quad (B8)$$

$$H_z(x,t) = \frac{nE_0}{Z_0} \left(\frac{n+1}{n^2+1}\right) \left[\exp\left(\frac{i2\pi nx}{\lambda_0}\right) + \left(\frac{n-1}{n+1}\right) \exp\left(-\frac{i2\pi nx}{\lambda_0}\right)\right] \exp(-i\omega t), \quad (B9)$$

$$\langle f(x,t) \rangle = \frac{2\pi n \varepsilon_0 E_0^2}{\lambda_0} \left(\frac{n^2-1}{n^2+1}\right)^2 \sin\left(\frac{4\pi nx}{\lambda_0}\right) \hat{x}. \quad (B10)$$

Finally, in the case of a single-layer antireflection coating on a substrate of refractive index $n_s = n^2$, the coating layer must have refractive index $n$ and thickness $d = \lambda_0/4n$. We find

$$E_y(x,t) = E_0 \frac{(n+1)}{2n} \left[\exp\left(\frac{i2\pi nx}{\lambda_0}\right) - \left(\frac{n-n^2}{n+n^2}\right) \exp\left(-\frac{i2\pi nx}{\lambda_0}\right)\right] \exp(-i\omega t), \quad (B11)$$

$$H_z(x,t) = \frac{nE_0}{Z_0} \frac{(n+1)}{2n} \left[\exp\left(\frac{i2\pi nx}{\lambda_0}\right) + \left(\frac{n-n^2}{n+n^2}\right) \exp\left(-\frac{i2\pi nx}{\lambda_0}\right)\right] \exp(-i\omega t), \quad (B12)$$

$$\langle f(x,t) \rangle = -\frac{\pi(n^2-1)^2 \varepsilon_0 E_0^2}{2n\lambda_0} \sin\left(\frac{4\pi nx}{\lambda_0}\right) \hat{x}. \quad (B13)$$

Alternatively, we can denote the substrate's refractive index by $n_s = n$, in which case the antireflection coating layer will have a refractive index $\sqrt{n}$ and thickness $d = \lambda_0/4\sqrt{n}$. The force-density of Eq.(B13) must then be written as follows:

$$\langle f(x,t) \rangle = -\frac{\pi(n-1)^2 \varepsilon_0 E_0^2}{2\sqrt{n}\lambda_0} \sin\left(\frac{4\pi\sqrt{n}x}{\lambda_0}\right) \hat{x}. \quad (B14)$$

Integrating the force-density of Eq.(B14) over the thickness $d$ of the layer now yields the total EM force acting on the antireflection coating as

$$\langle \boldsymbol{F}_0 \rangle = -\frac{(n-1)^2 \varepsilon_0 E_0^2}{4n} \hat{x}. \quad (B15)$$

The above expressions of EM force-density acting on various dielectric layers have been used in Sec.4 to arrive at the displacement function $u(x,t)$ under plane-wave illumination.

**Appendix C**

As pointed out in Sec.7.2, computation of $U_0(\omega)$ and $V_0(\omega)$ from initial conditions involves a subtlety. Consider, for instance, the following expansion of the initial condition $u(x, t = 0)$ in terms of the eigen-functions $u_\omega(x) = u'_\omega(x) + iu''_\omega(x)$ of an inhomogeneous medium:

$$u(x, t = 0) = \int_{-\infty}^{\infty} U_0(\omega') u_{\omega'}(x) d\omega'. \tag{C1}$$

The first step in determining $U_0(\omega)$ involves evaluating the (weighted) overlap integral between $u(x, t = 0)$ and $u_\omega(x)$ over the interval $[a, b]$, namely,

$$c(\omega) = \int_a^b \mu(x) u(x, t = 0) u_\omega(x) dx$$

$$= \int_{-\infty}^{\infty} U_0(\omega') \int_a^b \mu(x) u_\omega(x) u_{\omega'}(x) dx\, d\omega'$$

$$= U_0(\omega) \int_a^b \mu(x) u_\omega^2(x) dx + U_0(-\omega) \int_a^b \mu(x) u_\omega(x) u_\omega^*(x) dx$$

$$= U_0(\omega) \int_a^b \mu(x) [u'^2_\omega(x) - u''^2_\omega(x) + 2i u'_\omega(x) u''_\omega(x)] dx$$

$$+ U_0(-\omega) \int_a^b \mu(x) [u'^2_\omega(x) + u''^2_\omega(x)] dx. \tag{C2}$$

In the second step, we evaluate the (weighted) overlap integral between $u(x, t = 0)$ and $u_{-\omega}(x) = u_\omega^*(x)$ over the same interval $[a, b]$. Noting that $\mu(x)$ and $u(x, t = 0)$ are real-valued, we will have

$$c(-\omega) = c^*(\omega) = \int_a^b \mu(x) u(x, t = 0) u_\omega^*(x) dx$$

$$= \int_{-\infty}^{\infty} U_0(\omega') \int_a^b \mu(x) u_\omega^*(x) u_{\omega'}(x) dx\, d\omega'$$

$$= U_0(\omega) \int_a^b \mu(x) u_\omega^*(x) u_\omega(x) dx + U_0(-\omega) \int_a^b \mu(x) u_\omega^{*2}(x) dx$$

$$= U_0(\omega) \int_a^b \mu(x) [u'^2_\omega(x) + u''^2_\omega(x)] dx$$

$$+ U_0(-\omega) \int_a^b \mu(x) [u'^2_\omega(x) - u''^2_\omega(x) - 2i u'_\omega(x) u''_\omega(x)] dx. \tag{C3}$$

Combining Eqs.(C2) and (C3), we arrive at

$$\begin{bmatrix} \int_a^b \mu(x)[u'^2_\omega(x) - u''^2_\omega(x) + 2i u'_\omega(x) u''_\omega(x)] dx & \int_a^b \mu(x)[u'^2_\omega(x) + u''^2_\omega(x)] dx \\ \int_a^b \mu(x)[u'^2_\omega(x) + u''^2_\omega(x)] dx & \int_a^b \mu(x)[u'^2_\omega(x) - u''^2_\omega(x) - 2i u'_\omega(x) u''_\omega(x)] dx \end{bmatrix} \begin{bmatrix} U_0(\omega) \\ U_0(-\omega) \end{bmatrix} = \begin{bmatrix} c(\omega) \\ c^*(\omega) \end{bmatrix}. \tag{C4}$$

The values of $U_0(\omega)$ and $U_0(-\omega)$ may now be found by solving Eq.(C4), namely,

$$\begin{bmatrix} U_0(\omega) \\ U_0(-\omega) \end{bmatrix} = \frac{1}{\det}$$

$$\times \begin{bmatrix} \int_a^b \mu(x)[u'^2_\omega(x) - u''^2_\omega(x)] dx - 2i \int_a^b \mu(x) u'_\omega(x) u''_\omega(x) dx & -\int_a^b \mu(x)[u'^2_\omega(x) + u''^2_\omega(x)] dx \\ -\int_a^b \mu(x)[u'^2_\omega(x) + u''^2_\omega(x)] dx & \int_a^b \mu(x)[u'^2_\omega(x) - u''^2_\omega(x)] dx + 2i \int_a^b \mu(x) u'_\omega(x) u''_\omega(x) dx \end{bmatrix} \begin{bmatrix} c(\omega) \\ c^*(\omega) \end{bmatrix}, \tag{C5}$$

where the determinant (*det*) of the 2 × 2 matrix on the left-hand-side of Eq.(C4) is given by

$$det = \left\{\int_a^b \mu(x)[u_\omega'^2(x) - u_\omega''^2(x)]dx\right\}^2 + 4\left[\int_a^b \mu(x)u_\omega'(x)u_\omega''(x)dx\right]^2 - \left\{\int_a^b \mu(x)[u_\omega'^2(x) + u_\omega''^2(x)]dx\right\}^2$$

$$= 4\left[\int_a^b \mu(x)u_\omega'(x)u_\omega''(x)dx\right]^2 - 4\left[\int_a^b \mu(x)u_\omega'^2(x)dx\right]\left[\int_a^b \mu(x)u_\omega''^2(x)dx\right]. \quad (C6)$$

Clearly, $U_0(\omega)$ and $U_0(-\omega)$, which turn out to be a complex-conjugate pair, are uniquely determined via Eq.(C5) in terms of the overlap integral $c(\omega)$ defined in the first line of Eq.(C2) and the various integrals involving the real and imaginary parts of the eigen-function $u_\omega(x)$.